\documentclass[11pt]{article}

\usepackage[utf8]{inputenc}
\usepackage[T1]{fontenc}
\usepackage[margin=1in]{geometry}
\usepackage{times}

\usepackage{amsmath}
\usepackage{amssymb}
\usepackage{amsthm}
\usepackage{graphicx}
\usepackage{booktabs}
\usepackage{float}
\usepackage{caption}
\usepackage{enumitem}
\usepackage{xcolor}
\usepackage{url}

\usepackage{algorithm}
\usepackage{algorithmic}

\usepackage{tikz}
\usetikzlibrary{arrows.meta,positioning}

\usepackage{natbib}
\usepackage[hidelinks]{hyperref}

\theoremstyle{plain}
\newtheorem{theorem}{Theorem}
\newtheorem{lemma}[theorem]{Lemma}
\theoremstyle{definition}
\newtheorem{definition}{Definition}

\graphicspath{{figures/}}
\title{Collective-Influence-Guided Feedback-Vertex-Set Selection\\
       for Network Dismantling}

\author{Aryaman Ahluwalia\\
  School of Computer Science\\
  University of Windsor\\
  \texttt{aryamanahluwalia012@gmail.com}
  \and
  Asish Mukhopadhyay\\
  School of Computer Science\\
  University of Windsor\\
  \texttt{asishm@uwindsor.ca}}

\date{\today}

\begin{document}

\maketitle

\begin{abstract}
Network dismantling asks for the smallest node set whose removal leaves no
connected component larger than a chosen threshold. It is NP-hard, and the
heuristics used in practice report margins over one another of a percentage
point or two, small enough that attributing a result to the rule that produced
it becomes the difficulty. We build a two-phase structural pipeline and use it
to make that attribution measurable. Phase~1 decycles the 2-core by
depth-first-search back-edge detection, removing one vertex of each certified
cycle by its radius-one Collective Influence; Phase~2 fragments the resulting
forest by greedy Collective Influence removal. Phase~1 returns a valid feedback
vertex set in $O(|F|(n+m))$ time, and no approximation bound is claimed for the
pipeline. Against Random removal, CoreHD, Collective Influence and
WEAK-NEIGHBOR on Erd\H{o}s--R\'enyi and configuration-model graphs and two real
contact networks, the pipeline is competitive with the established heuristics
but does not beat them: its best variant is level with WEAK-NEIGHBOR at $0.191$
on Erd\H{o}s--R\'enyi graphs and ahead of CoreHD's $0.193$, a margin of $0.2$
points well inside instance-to-instance spread, and it places first on none of
the four networks. Three controls explain what that comparison reports. Every
decycling strategy hands off to the same greedy fill once the graph becomes
acyclic, and that shared tail is $84$ to $86\%$ of each synthetic sequence at
every size from $n=500$ to $n=4000$. Holding the second phase fixed and optimal
shows that $86$ to $90\%$ of the total removal budget is the feedback vertex
set itself. Separating the global cycle ordering from the influence score shows
that the ordering carries the gain and the score is not separable from degree
on one family and worse than it on the other. A cycle-rank bound confines
competitive performance to sparse graphs. The contribution is empirical and
explanatory.
\end{abstract}

\section{Introduction}
\label{sec:introduction}

\subsection{Motivation}
\label{sec:motivation}

When an infection spreads through a population, it does not travel through the
population at large but along a network of contacts: who is close to whom, who
meets whom, which individuals bridge otherwise separate groups. The COVID-19
pandemic made this visible on a global scale, in that the course of an outbreak
depends not on the pathogen alone but on the shape of the contact network it
moves through. The interventions governments reached for, such as distancing,
closures and deciding whom to vaccinate first, were all attempts to change that
structure, breaking the connectivity along which transmission travels. The
pandemic also exposed the cost of acting on that structure without understanding
it. Blunt, population-wide measures were expensive and imperfectly targeted, and
the question of \emph{which} individuals or connections mattered most to the
spread, and \emph{why}, was rarely one that could be answered transparently.

That question is the subject of this paper, in a form abstracted from any single
disease. A network sustains large-scale spreading precisely when it contains a
giant connected component: a single cluster spanning a constant fraction of the
nodes, through which a contagion can reach almost everyone. Remove the right
nodes and that component shatters into pieces too small to sustain an outbreak;
remove the wrong ones and it barely changes. The problem of choosing those
nodes, a minimum-size set whose removal fragments the network below a chosen
scale, is known as \emph{network dismantling}, and it is the same problem as
targeted immunisation read in the other direction: the nodes whose removal most
fragments the contact network are the nodes whose vaccination most impedes the
epidemic. What is learned about one is directly a statement about the other.

\subsection{The Problem and Its Difficulty}
\label{sec:problem}

Formally, network dismantling asks for the smallest set of nodes whose removal
leaves no connected component larger than a prescribed threshold. The problem is
NP-hard, and on the large networks where it matters, such as social graphs,
contact networks and infrastructure, exact solution is out of reach. Practical
work therefore turns to heuristics: rules that decide which node to remove next.
Two families dominate the current literature. Combinatorial heuristics such as
CoreHD and Collective Influence rank nodes by a structural score and remove them
greedily; learning-based methods such as graph neural network solvers train a
model to make those decisions. Both achieve strong empirical performance.

What neither makes easy is \emph{attribution}. The margins separating these
methods are routinely a percentage point or two, and a dismantling procedure is
a sequence of many decisions, so a reported advantage is difficult to trace back
to the rule that is supposed to have produced it. For the immunisation reading
of the problem, where a recommendation may translate into who is protected
first, knowing which structural decision did the work is not a secondary
concern. The approach taken here is therefore to build a dismantling procedure
out of steps that can be examined one at a time, and to measure how much of the
procedure's performance each step is responsible for.

\subsection{Approach and Contributions}
\label{sec:contributions}

This paper proposes, evaluates, and explores a two-phase structural pipeline.
The first phase removes a feedback vertex set, a set of nodes whose deletion
leaves the network acyclic, using a depth-first-search back-edge heuristic, on
the reasoning that the cycles a network contains are what make its connectivity
robust. The second phase fragments the resulting forest by greedy removal ranked
by Collective Influence.

The contribution is empirical and explanatory. The findings are the following.

\begin{itemize}[itemsep=0.15\baselineskip]
  \item Three properties are established for the procedure: the first phase
    returns a valid feedback vertex set, the procedure terminates, and it runs
    in polynomial time. Its performance is settled by measurement.
  \item On sparse synthetic graphs the method is competitive with the
    established heuristics, at margins that lie inside instance-to-instance
    spread. It places first on none of the four networks evaluated.
  \item Most of each synthetic removal sequence is produced by a phase shared
    across every variant, so the rankings there reflect the residual each method
    leaves at handoff more than the decisions it continues to make. On the
    real-world networks the sequences are genuinely method-driven, and the
    comparison is correspondingly sharper.
  \item Holding that shared phase fixed and optimal isolates the decycling step,
    and shows that the great majority of the removal budget is the feedback
    vertex set itself. Fragmentation is not where dismantling is decided.
  \item Within the decycling step, the gain belongs to the global cycle ordering
    rather than to the influence score that motivated it. Separating the two
    shows the score is not distinguishable from degree on one synthetic family
    and worse than degree on the other.
  \item A density constraint intrinsic to decycling-based dismantling confines
    competitive performance to sparse graphs. This is a property of the family
    of approaches rather than of any one implementation, and we characterise the
    resulting envelope rather than averaging over it.
\end{itemize}

What the pipeline delivers is a measured account of how decycling-first
dismantling behaves, what each of its phases contributes, and where its limits
lie.

\subsection{Organisation}
\label{sec:organisation}

Section~\ref{sec:related-work} reviews the network dismantling literature and
the structural and algorithmic background the pipeline draws on.
Section~\ref{sec:preliminaries} fixes notation and states the structural results
the pipeline's correctness argument uses. Section~\ref{sec:method} develops the
two phases, the three selection rules built on the first, and their cost.
Section~\ref{sec:theory} states what is and is not established, and to which
subroutine each bound belongs. Section~\ref{sec:experiments} evaluates the
pipeline against established baselines on synthetic and real networks and
isolates the contribution of each phase. Section~\ref{sec:discussion} reports
the controls that qualify those results and establishes the density and size
envelope within which the method is competitive at all.
Section~\ref{sec:conclusion} summarises what the measurements support and sets
out the questions the work leaves open.

\section{Related Work}
\label{sec:related-work}

\subsection{The Network Dismantling Problem}

Network dismantling, sometimes referred to in the literature as optimal
percolation or network fragmentation, asks for the smallest set of nodes whose
removal breaks a network into components no larger than some prescribed
threshold, typically expressed as a small fraction of the original network
size~\citep{braunstein2016}. The problem is of direct practical relevance to
epidemic control, where removing a set of individuals (through vaccination or
quarantine) is intended to halt the spread of a pathogen; to infrastructure and
cybersecurity, where an adversary or defender seeks to identify critical nodes
whose failure or removal disconnects a system; and to the containment of
misinformation or influence cascades on social networks~\citep{artime2024}.
Formally, the problem is closely related to, but distinct from, the classical
feedback vertex set (FVS) problem: while FVS asks for a minimum set of nodes
intersecting every cycle in a graph, dismantling additionally requires that the
\emph{residual forest itself} be broken into sufficiently small tree fragments,
since a large acyclic component can still support significant residual
connectivity. Both problems are NP-hard on general graphs, which has driven the
network science community toward heuristic, spectral, and more recently
learning-based approximations.

\subsection{Percolation-Based and Collective Influence Approaches}

A foundational contribution to this area is the optimal percolation framework
of \citet{morone2015}, which reframes network dismantling as an optimization
problem over the percolation threshold of the network and introduces Collective
Influence (CI), a centrality measure derived from the theory of spin-glasses on
locally tree-like graphs, as a principled criterion for ranking nodes by their
structural importance in maintaining connectivity. CI scores a node by
aggregating the degree contributions of nodes at a fixed distance $\ell$ from
it, and has been shown to substantially outperform classical centrality measures
such as degree, betweenness, and eigenvector centrality when used to greedily
remove nodes. A key theoretical property of CI, relevant to the present work, is
that it is only an approximation on graphs containing cycles, but becomes an
\emph{exact} measure of a node's structural importance on trees, where the
locally tree-like assumption underlying its derivation holds without
approximation error. Section~\ref{sec:ci-trees} returns to this property and
fixes its scope: exactness belongs to the score rather than to the greedy policy
built on it, and at the radius of one actually used here the tree condition asks
very little.

\citet{braunstein2016} extended this line of work by explicitly recasting
network dismantling as a staged procedure. Their algorithm has three parts:
decycling the network by computing a minimal feedback vertex set using a
message-passing (Min-Sum) algorithm derived from statistical physics; breaking
the resulting forest into components below the target size; and finally
reinserting, greedily, those removed nodes that can be returned without pushing
the largest component back over the threshold. The second of these stages is the
one part of the dismantling problem that admits an exact polynomial solution:
breaking a forest into components of size at most $\tau$ with the fewest
deletions can be solved by dynamic programming in $O(n\tau^{2})$ time, although
they note that a greedy alternative is cheaper and performs almost as well in
practice. This decomposition, decycle first and then fragment the residual
forest, established the conceptual template that much of the subsequent
dismantling literature, including the present work, continues to build upon.
Their exact forest-breaking algorithm is used in
Section~\ref{sec:fixed-completion} as an experimental control, and their
reinsertion stage is the subject of the reappraisal by
\citet{fan2020reinsertion} discussed in Section~\ref{sec:positioning}.

\citet{li2021longloop} refine the premise underlying this template. Loop-focused
decycling treats every cycle as equally worth breaking, but the short loops
inside a densely connected local cluster contribute little to the long-range
connectivity that dismantling targets, whereas the long loops joining separate
clusters carry it. They accordingly propose a bipartite factor-graph formulation
that retains the long loops while collapsing each dense local cluster into a
single factor node, develop a mean-field theory for the resulting long-loop
feedback vertex set problem, and use its message-passing equations as a
dismantling solver, reporting improved performance on real networks. The
observation that decycling can spend removals on cycles whose destruction does
not serve the dismantling objective recurs throughout this paper.

\citet{ren2019} generalized this framework further by formulating network
dismantling as a problem over tree decompositions, unifying node removal and
edge removal formulations and demonstrating that this generalized view yields
improved dismantling sets across a broad range of network topologies. Their work
reinforced the value of exploiting tree-like structure in the residual graph
after decycling, a theme central to the fragmentation phase employed here.

\subsection{Heuristic Approaches: CoreHD}

CoreHD~\citep{zdeborova2016} is a widely used heuristic baseline in the
dismantling literature. It operates by iteratively computing the $k$-core of the
residual network and greedily removing the highest-degree node within the core,
repeating this process until the network is sufficiently fragmented. CoreHD is
computationally efficient and consistently competitive with more sophisticated
methods, particularly on sparse networks, which has made it a standard point of
comparison for newly proposed dismantling algorithms. Its behaviour on random
graph ensembles has been characterised analytically: \citet{schmidt2019} analyse
CoreHD on the configuration model via a set of deterministic differential
equations, and use that analysis to derive upper bounds on the size of the
\emph{minimal contagious set}, the smallest vertex set whose removal empties the
$k$-core, which improve on previously known bounds. The bound so obtained is a
statement about the optimum for the ensemble, established by way of CoreHD's
analysed performance, rather than a statement about what CoreHD returns for a
given input graph. Results of this kind apply to the ensemble rather than to any
particular empirical network, and we accordingly treat CoreHD as an empirical
baseline throughout.

\subsection{Learning-Based Approaches}

More recently, the dismantling problem has been approached through deep
learning~\citep{grassia2021}, most notably by FINDER~\citep{fan2020}, which
trains a graph neural network via reinforcement learning to sequentially select
nodes for removal, learning a removal policy directly from examples of small
synthetic graphs and generalizing it to much larger real-world networks.
Learning-based methods of this kind can achieve state-of-the-art empirical
performance and have the appealing property of adapting to structural
regularities the designer may not have anticipated. Their principal drawback is
a dependence on the training distribution: performance on a given network is
contingent on how far that network resembles the graphs the policy was trained
on, and the method offers no way to check that resemblance in advance.

\subsection{Feedback Vertex Set Approximation Algorithms}

Independently of the dismantling literature, the feedback vertex set problem has
a long history within approximation algorithms research. Classical results
establish 2-approximation algorithms for FVS on general graphs via the
local-ratio technique and via LP-rounding, guaranteeing a decycling set of size
at most twice the optimum. These algorithms, in contrast to the message-passing
and heuristic decycling procedures used elsewhere in the dismantling literature,
come with a formal worst-case bound on solution quality. Such a bound is often
assumed to be paid for in practice, by returning larger feedback vertex sets
than heuristics tuned to the dismantling objective.
Section~\ref{sec:decycling-bound} measures that trade-off on the graphs used
here and does not find it: the local-ratio algorithm returns a set within half a
percentage point of CoreHD's and dismantles competitively. What its bound does
not do is reach any dismantling quantity, which is a separate matter taken up
there.

\subsection{Positioning of This Work}
\label{sec:positioning}

The literature surveyed above has a shape worth naming. A great many methods
have been proposed, most of them share the decycle-then-fragment template of
\citet{braunstein2016}, and the margins separating them are routinely one or two
percentage points. Two strands of prior work have already asked what those
margins mean. \citet{wandelt2018} re-implemented thirteen methods across twelve
families of random graph, on the ground that results published across different
venues on different datasets could not be compared as they stood.
\citet{fan2020reinsertion} went further and isolated a single shared component,
the greedy reinsertion step that most published pipelines apply at the end,
reporting that it rather than the selection rule largely determines the final
ranking.

This paper takes up that question for the decycling-first template in
particular. It builds a two-phase pipeline, decycling the 2-core by
depth-first-search back-edge detection and fragmenting the residual forest by
greedy radius-one Collective Influence, and then uses that pipeline as an
instrument: what does each phase actually contribute, how much of a reported
dismantling result is attributable to the rule under study, and where does the
template stop being competitive at all.

The pipeline is a method and is benchmarked as one. Three properties are
established for it, that its first phase returns a valid feedback vertex set,
that it terminates, and that it runs in polynomial time, and
Section~\ref{sec:experiments} places it against Random removal, CoreHD,
Collective Influence, WEAK-NEIGHBOR and the local-ratio decycler of
\citet{bafna1999} on synthetic and real networks. It is competitive with these
on sparse graphs. It is not, however, offered as a replacement for them.

What this work contributes is the measurement. Section~\ref{sec:discussion}
quantifies how much of each removal sequence is produced by a phase shared
across every variant, Section~\ref{sec:fixed-completion} holds that phase fixed
and optimal in order to isolate the decycling step, and
Section~\ref{sec:density} establishes the density range over which
decycling-first dismantling remains competitive at all. The account that results
is specific to this family of methods, and it is the reason the pipeline was
built.

\section{Preliminaries}
\label{sec:preliminaries}

Nothing in this section is new. It is collected here because the pipeline's
correctness argument, its termination test and the density limit of
Section~\ref{sec:density} all depend on it. Graph-theoretic definitions follow
standard usage~\citep{newman2018}.

Let $G = (V, E)$ be a finite, simple, undirected graph with $n = |V|$ vertices
and $m = |E|$ edges. For a vertex $v \in V$, $d(v)$ denotes its degree, and
$\Delta$ denotes the maximum degree of $G$. For a subset $S \subseteq V$, $G[S]$
denotes the subgraph of $G$ induced by $S$. The distance $\mathrm{dist}(u,v)$
between two vertices is the number of edges on a shortest $u$--$v$ path.

\subsection{Cores, Cycles and Feedback Vertex Sets}

\begin{definition}[$k$-core and coreness]
\label{def:kcore}
The \emph{$k$-core} of $G$ is the maximal induced subgraph of $G$ in which every
vertex has degree at least $k$. The \emph{coreness} $\mathrm{core}(v)$ of a
vertex $v$ is the largest $k$ such that $v$ belongs to the $k$-core.
\end{definition}

The definition relaxes that of a clique. A clique asks every vertex to be
adjacent to every other; a $k$-core asks only that each vertex keep $k$
neighbours inside the subgraph, which is a far weaker demand and is why
$k$-cores are large where cliques are tiny. Two properties are worth fixing now,
because both are used later. The $k$-core need \emph{not} be connected: nothing
in the definition ties its pieces together, and a graph made of two far-apart
triangles has a 2-core with two components. And it is unique, which is what
``maximal'' delivers: peeling vertices of degree below $k$ in any order reaches
the same subgraph, since a vertex that cannot survive is never rescued by
deleting others.

\begin{definition}[Cycle rank]
\label{def:cycle-rank}
The \emph{cycle rank} (or circuit rank) of $G$ is
\[
\mu(G) \;=\; m - n + c,
\]
where $c$ is the number of connected components of $G$.
\end{definition}

The formula is easiest to read off a spanning forest. Take a spanning tree of
each of the $c$ components of $G$ and call the result $T$. A tree on $n_i$
vertices has $n_i - 1$ edges, so $T$ has $n - c$ edges altogether, and $G$ has
$m - (n-c) = \mu(G)$ edges that $T$ leaves out. Now put one of those omitted
edges back. Its endpoints $u$ and $v$ are already joined by a unique path in
$T$, so adding $e = \{u,v\}$ creates exactly one cycle. The $\mu(G)$ cycles
built this way are the \emph{fundamental cycles} of $T$. They are independent,
and every cycle of $G$ is a combination of some of them under symmetric
difference. So $\mu(G)$ counts at once the minimum number of \emph{edges} whose
deletion leaves a forest, the number of independent cycles of $G$, and the
dimension of the cycle space of $G$ over $\mathrm{GF}(2)$. The count does not
depend on which spanning forest was chosen, and $\mu(G) = 0$ exactly when $G$ is
a forest.

\begin{definition}[Feedback vertex set]
\label{def:fvs}
A \emph{feedback vertex set} (FVS), equivalently a \emph{decycling set}, of $G$
is a subset $F \subseteq V$ such that $G[V \setminus F]$ is acyclic, that is, a
forest. The minimum cardinality of such a set is the \emph{decycling number} of
$G$, written $\nabla(G)$.
\end{definition}

Cycle rank matters throughout this paper, but its role is easy to state wrongly.
It is not a lower bound on the size of a decycling set. Take $K_4$:
$\mu(K_4) = 3$, but deleting any two vertices leaves a single edge, so two are
enough. A single vertex can sit on several independent cycles, so a decycling
set can be smaller than the number of cycles it destroys. What $\mu$ does
control is the edge surplus $m - n$, which by Definition~\ref{def:cycle-rank}
equals $\mu(G) - c$. That surplus, together with a bound on the maximum degree,
is what forces a decycling set to be large.

\begin{lemma}
\label{lem:fvs-density-bound}
Let $G$ have maximum degree $\Delta \ge 2$, and let $F$ be any feedback vertex
set of $G$. Then
\[
|F| \;>\; \frac{m - n}{\Delta - 1} \;=\; \frac{\mu(G) - c}{\Delta - 1}.
\]
\end{lemma}

\noindent\emph{Proof.} Assume $F \ne V$; the case $F = V$ is immediate, since
$m \le n\Delta/2$ makes the right-hand side smaller than $n = |F|$. Write
$H = G[V \setminus F]$, a graph on $n - |F| \ge 1$ vertices, and bound $|E(H)|$
from both sides. Above: $H$ is acyclic, so it is a forest and has at most
$n - |F| - 1$ edges. Below: deleting $F$ destroys only edges with an endpoint in
$F$, of which there are at most $\sum_{v \in F} d(v) \le |F|\Delta$, so
$|E(H)| \ge m - |F|\Delta$. Chaining the two gives $m - |F|\Delta < n - |F|$,
hence $m - n < |F|(\Delta-1)$, and dividing by $\Delta - 1 > 0$ yields the first
form. Definition~\ref{def:cycle-rank} rearranged gives $m - n = \mu(G) - c$ and
hence the second. \hfill$\square$

Rewriting the surplus in terms of the average degree
$\langle k \rangle = 2m/n$ gives the form used repeatedly below. When
$\langle k \rangle \ge 4$ we have $m \ge 2n$, so $m - n \ge n$, and
Lemma~\ref{lem:fvs-density-bound} becomes
\[
|F| \;>\; \frac{n}{\Delta - 1}.
\]
If the maximum degree stays bounded as the graph grows, $n/(\Delta-1)$ is a
fixed fraction of the whole network. The bound holds for every valid feedback
vertex set, not only for the one some algorithm happens to return, so no choice
of method avoids it. Dividing by $n$ puts it in the units the tables use, which
is why Table~\ref{tab:networks} carries a $\mu/n$ column, and
Section~\ref{sec:density} is where the bound does its work.

Definition~\ref{def:fvs} has an equivalent covering form used just as often: $F$
is a feedback vertex set if and only if it contains at least one vertex of every
cycle of $G$. Both directions are immediate. A cycle avoiding $F$ entirely would
keep all of its vertices and edges in $G[V \setminus F]$, which would then not be
acyclic; and any cycle of $G[V \setminus F]$ is a cycle of $G$ meeting no vertex
of $F$. Phase~1 works in the covering form, finding cycles and paying for each by
committing a vertex. That form also shows why the problem resists a direct
attack: the number of cycles in a graph can grow exponentially with $n$, so no
method can list them and cover them one by one.

Two further points about feedback vertex sets matter for
Section~\ref{sec:decycling-bound}. The first is the difference between
\emph{minimum} and \emph{minimal}. A feedback vertex set is minimum if none of
smaller cardinality exists. It is \emph{inclusion-minimal} if no proper subset of
it is a feedback vertex set, which is the same as saying that every $v \in F$
lies on a cycle that $F \setminus \{v\}$ misses. Minimum implies
inclusion-minimal, but the converse can fail badly. Let $G_k$ consist of a
shared vertex $h$ and $k$ triangles $\{h, a_i, b_i\}$ meeting only at $h$.
Deleting $h$ leaves $k$ disjoint edges, so $\nabla(G_k) = 1$. The set
$\{a_1, \dots, a_k\}$ is also a feedback vertex set, since what remains is a
star, and it is inclusion-minimal, since dropping any $a_j$ restores the
triangle $\{h, a_j, b_j\}$. Its size is $k$. Minimality on its own therefore
says nothing about size, and in Section~\ref{sec:decycling-bound} it is one
hypothesis among several rather than the source of the result.

The second point is that computing $\nabla(G)$ is NP-hard. The decision version
of feedback vertex set is one of Karp's original twenty-one NP-complete
problems~\citep{karp1972}. No Phase-1 procedure in this paper therefore returns
a set that is certainly smallest, and at the sizes tested here $\nabla(G)$
cannot be computed either, so how far off any of them are cannot be measured.

\subsection{Dismantling, and How It Is Scored}

\begin{definition}[Network dismantling]
\label{def:dismantling}
Given a threshold $\tau$, the \emph{network dismantling problem} asks for a
minimum-cardinality set $S \subseteq V$ such that every connected component of
$G[V \setminus S]$ has at most $\tau$ vertices.
\end{definition}

The threshold is what gives Definition~\ref{def:dismantling} its content. The
criterion caps component \emph{size}; it does not ask for the residual graph to
be disconnected. Requiring only disconnection would be a weak test, since one
cut vertex often achieves it while leaving two enormous pieces behind. A network
whose largest surviving component holds one per cent of its vertices, by
contrast, has lost whatever capacity motivated attacking it, whether that was
spreading an epidemic or carrying traffic. We therefore take $\tau$ as a
fraction of $n$ rather than as an absolute constant, so that it means the same
thing on networks of different sizes.

The problem is NP-hard even at $\tau = 1$, where it reduces to minimum vertex
cover: every component of $G[V \setminus S]$ is a single vertex exactly when no
edge survives, so $S$ must be a vertex cover~\citep{karp1972}. Heuristics are
not a convenience here; at these scales they are the only option.

Dismantling and decycling are also distinct problems, though closely related
ones. A feedback vertex set leaves a forest, and a forest can still have large
components, so decycling by itself does not dismantle. \citet{braunstein2016}
show that on random graphs the two problems need roughly the same number of
removals, but that this correspondence fails for heavy-tailed degree
distributions. The two-phase architecture studied here sits in that gap: Phase~1
solves the decycling problem, and Phase~2 exists because solving it was not
enough.

The evaluation does not measure the size of a set meeting a fixed $\tau$. Every
method studied here produces an \emph{ordering} of the vertices rather than a
set, so the natural way to summarise it is to watch the largest connected
component (LCC) shrink as vertices are removed in that order. That curve reports
the outcome at every threshold at once, instead of at one threshold chosen in
advance. Here $f$ denotes a fraction of the network, so $fn$ is that fraction of
the $n$ vertices. Write $g(f)$ for the size of the LCC once those $fn$ vertices
have been deleted, again as a fraction of $n$. It starts at the LCC fraction of
the intact graph, decreases as $f$ grows, and reaches $0$ at $f = 1$. Two
summaries of $g$ are reported throughout.

The \emph{collapse fraction}
\[
f^{*} \;=\; \min \{\, f : g(f) < 0.05 \,\}
\]
is the fraction of vertices that must be removed before the LCC first falls
below five per cent of the network. It is Definition~\ref{def:dismantling} at
$\tau = 0.05n$, with one restriction: the set has to be built by following the
method's own order and stopping somewhere, instead of being chosen freely.
Stopping at $f^{*}n$ gives a set that meets the threshold, so
$f^{*}n \ge |S_{\mathrm{opt}}|$. The difference between the two is the combined
cost of using a heuristic and of being tied to that order.

The \emph{area under the curve},
\[
\mathrm{AUC} \;=\; \int_{0}^{1} g(f)\,\mathrm{d}f,
\]
is the mean LCC fraction across all removal budgets. It complements $f^{*}$ by
rewarding a sequence that shrinks the LCC steadily instead of one that does
nothing until a late and abrupt collapse. For scale when reading the tables: a
sequence that never reduces the LCC scores near $1$, one that shatters the
network immediately scores near $0$, and random removal on the sparse synthetic
graphs of Section~\ref{sec:setup} scores about $0.35$. Lower is better for both
quantities.

Finally, some notation for trees. The \emph{eccentricity} of a vertex of a tree
$T$ is $\epsilon(v) = \max_{u} \mathrm{dist}(u,v)$. A \emph{center} of $T$ is a
vertex of minimum eccentricity; a \emph{centroid} is a vertex minimizing the
size of the largest component of $T - v$. One optimises distance and the other
optimises size, and they are usually not the same vertex.
Section~\ref{sec:centroid} takes up what that difference is worth.

\subsection{Reduction to the 2-Core}
\label{sec:2core}

Phase~1 begins by reducing the input graph to its 2-core: delete vertices of
degree at most one, repeatedly, until none remain. Two properties justify the
reduction, and the termination test of the decycler in
Section~\ref{sec:selection} depends on the first of them.

Pruning in different orders cannot produce different 2-cores. No vertex loses a
neighbour by being placed in a larger graph, so if $G[A]$ and $G[B]$ both have
minimum degree at least $2$, so does $G[A \cup B]$; the union of every vertex set
with that property therefore has it too, and is the unique largest such set,
which is the 2-core of Definition~\ref{def:kcore} at $k = 2$. Any pruning order
that terminates stops at a subgraph of minimum degree at least $2$, and the
argument below shows it never deletes a vertex the 2-core contains, so all
orders agree.

\begin{lemma}
\label{lem:cycles-in-2core}
Every cycle of $G$ lies entirely within the 2-core of $G$. Consequently $G$ is a
forest if and only if its 2-core is empty.
\end{lemma}

\noindent\emph{Proof.} Let $v_1, v_2, \dots, v_t$ be the vertices in the order
the pruning deletes them, let $G_0 = G$, and let $G_i$ be the graph remaining
after $v_1, \dots, v_i$ have been deleted, so that $G_t$ is the 2-core. We show
by induction on $i$ that no vertex lying on a cycle of $G$ is ever deleted.
Suppose the claim holds for all $j < i$, and suppose for contradiction that
$v_i$ lies on a cycle $Z$ of $G$. The vertex $v_i$ has two distinct neighbours on
$Z$, say $x$ and $y$. Both lie on the cycle $Z$, so by the induction hypothesis
neither is among $v_1, \dots, v_{i-1}$; both are therefore still present in
$G_{i-1}$, along with the edges $v_ix$ and $v_iy$, since an edge is lost only
when one of its endpoints is deleted. Hence $\deg_{G_{i-1}}(v_i) \ge 2$, and the
pruning rule, which deletes only vertices of current degree at most one, cannot
have selected $v_i$ at step $i$. This contradiction establishes the induction.

No vertex of any cycle of $G$ is therefore deleted, so $V(Z) \subseteq V(G_t)$
for every cycle $Z$, and since $G_t$ is an induced subgraph it retains every edge
of $Z$ as well; $Z$ is thus a cycle of the 2-core. For the consequence: if $G$ is
a forest it has no cycles, and pruning a finite forest deletes every vertex,
since a non-empty forest always has a leaf or an isolated vertex, so the 2-core
is empty. Conversely, if $G$ has a cycle then that cycle's vertices survive into
the 2-core, which is therefore non-empty. \hfill$\square$

\begin{lemma}
\label{lem:fvs-2core}
Let $C$ denote the vertex set of the 2-core of $G$. Then $F \subseteq V$ is a
minimum feedback vertex set of $G$ if and only if $F$ is a minimum feedback
vertex set of $G[C]$. In particular $\nabla(G) = \nabla(G[C])$.
\end{lemma}

\noindent\emph{Proof.} By Lemma~\ref{lem:cycles-in-2core}, $G$ and $G[C]$ have
the same cycles, so by the covering form of Definition~\ref{def:fvs} a set is a
feedback vertex set of one exactly when it covers that common family. Any
minimum feedback vertex set lies inside $C$: a vertex outside it sits on no cycle
and could be dropped, contradicting minimality. Minimum sets of the two graphs
are therefore drawn from the same collection of candidates and measured by the
same cardinality, so they coincide. \hfill$\square$

Neither result is new. They are proved
here because the pipeline's correctness argument uses them, and because the
literature states the second only in the weaker form it needs. The $k$-core is
due to~\citet{seidman1983}, and the peeling computation is the standard
linear-time core decomposition of~\citet{batagelj2011}. The equivalence in
Lemma~\ref{lem:cycles-in-2core} between acyclicity and an empty 2-core is well
known, and the dismantling literature applies it directly without setting out a
proof; the CoreHD decycler of~\citet{zdeborova2016} confines every removal to the
2-core on exactly this ground and stops once that core is empty.
Lemma~\ref{lem:fvs-2core} adds the statement in terms of \emph{minimum} feedback
vertex sets, which is what makes the reduction lossless and not just convenient.

The reduction runs in $O(m)$ time, since each vertex is deleted at most once and
each deletion touches only the adjacency list of the deleted vertex. It also
gives a measurement before any dismantling begins. The fraction of vertices that
survive into the 2-core is the fraction of the graph lying on cycles, and
therefore an advance estimate of how much work Phase~1 faces. On the sparse
synthetic families the 2-core keeps about three quarters of the vertices,
$74.9\%$ on Erd\H{o}s--R\'enyi and $75.6\%$ on the configuration model at
$n = 500$ and $\langle k \rangle = 3$. On the two real contact networks it
removes almost nothing, leaving $97.3\%$ of Add Health and all of the CollegeMsg
proxy.

\section{The Pipeline}
\label{sec:method}

The pipeline splits network dismantling into two sequential phases. Phase~1
enforces \emph{acyclicity}: it computes and removes a feedback vertex set on the
2-core of the input graph, leaving a forest. Phase~2 performs
\emph{fragmentation}: it breaks up that residual forest by greedy removal. The
\emph{handoff}, the fraction of vertices already removed when the graph first
becomes a forest, separates the cost of decycling from the cost of fragmenting,
and is reported for every method in this paper.

\subsection{Design Requirements from a First Realisation}
\label{sec:requirements}

The design below is the second realisation of that architecture. The first,
which is reported in full elsewhere and appears here only as an ablation
baseline, paired a local-ratio feedback vertex set procedure with recursive tree
center removal, and failed in one specific way per phase. The two failures are
not of the same kind, and the distinction decides what can be done about each.

Phase~1 never made a choice. The two-phase architecture invites the reading that
Phase~1 repeatedly finds a cycle and chooses one of its vertices to delete, so
that the quality of the set depends on how that choice is made. The implemented
local-ratio procedure did not work this way. Its semi-disjoint-cycle branch, the
only branch that commits a single chosen vertex, fired three times across forty
benchmark instances at $n = 500$, twice on one Erd\H{o}s--R\'enyi graph, once on
another, and on none of the twenty configuration-model graphs, against a mean of
$2.0$ tight-set passes per instance. Virtually the entire set therefore came
from the tight-set branch, which commits every member of a tight set at once and
so makes no selection among vertices at all. Without the reverse-delete pass
those few passes absorbed almost the whole 2-core, committing $72.9\%$ and
$72.2\%$ of the network on the two families against inclusion-minimal sets of
$21.7\%$ and $20.9\%$, an inflation of $3.36\times$ and $3.46\times$.

Phase~2, by contrast, did choose, and chose badly. Removing the first tree
center in component-enumeration order consults neither vertex importance nor
component size, and in consequence attacked the largest component in fewer than
one removal in fifteen while deleting vertices of mean degree below $0.5$.

Each failure admits only one kind of repair. A procedure that never chooses
cannot be improved by giving it a better criterion, because there is no point at
which a criterion would be consulted: no ranking of a tight set's vertices would
change what it removes. Phase~1's procedure is therefore replaced outright, by a
DFS back-edge scheme that deletes one vertex per detected cycle and so creates
the per-cycle decision an importance measure can govern
(Section~\ref{sec:selection}). Phase~2 requires no such replacement. It already
removes one vertex at a time and already consults a criterion, so it is the
criterion that is exchanged: the geometric center gives way to Collective
Influence (Section~\ref{sec:phase2}). The two-phase decomposition and the 2-core
reduction are kept unchanged.

\subsection{Phase 1: Importance-Aware Cycle-Node Selection}
\label{sec:selection}

Phase~1 uses \emph{DFS back-edge detection}: a depth-first traversal is run on
the 2-core; each back edge $(u,v)$ certifies a cycle, namely the tree path from
$v$ to $u$ closed by the edge itself; an importance score decides where that
cycle is broken; the residual graph is re-pruned to its 2-core; and the process
repeats until the graph is a forest. The traversal mechanics are those of
Algorithm~\ref{alg:back-edge}; what changes between the three strategies below is
only the choice of \emph{which} back edge is processed and \emph{which} vertex of
its cycle is removed. Because exactly one vertex is deleted per iteration, this
procedure makes a genuine per-cycle choice.

\begin{algorithm}[htbp]
\caption{DFS back-edge cycle elimination (the Phase-1 procedure)}
\label{alg:back-edge}
\begin{algorithmic}[1]
\REQUIRE Graph $G$, selection rule $\sigma$
\ENSURE Removal list $R$ such that $G - R$ is a forest
\STATE $H \leftarrow G$; $R \leftarrow$ empty list
\LOOP
    \STATE $C \leftarrow$ vertex set of the 2-core of $H$
    \IF{$C = \emptyset$ \OR $H[C]$ is acyclic}
        \STATE \textbf{break}
    \ENDIF
    \STATE $B \leftarrow \textsc{FindBackEdges}(H, C)$
    \STATE $v \leftarrow \sigma(H, C, B)$ \COMMENT{the rule of Section~\ref{sec:baseline-dfs}, \ref{sec:option1} or \ref{sec:option2}}
    \STATE append $v$ to $R$; delete $v$ from $H$
\ENDLOOP
\RETURN $R$
\vspace{0.6em}
\STATE \textbf{procedure} \textsc{FindBackEdges}($H$, $C$)
\STATE $B \leftarrow$ empty list; $\mathrm{visited} \leftarrow \emptyset$; $t \leftarrow 0$
\FOR{each $s \in C$, in ascending label order}
    \IF{$s \in \mathrm{visited}$}
        \STATE \textbf{continue}
    \ENDIF
    \STATE $\mathrm{visited}[s] \leftarrow t$; $t \leftarrow t+1$; push $(s, \textbf{nil})$ onto an explicit stack
    \WHILE{the stack is not empty}
        \STATE $(v, p) \leftarrow$ top of stack
        \IF{$v$ has an unexplored neighbour $u$ in $H[C]$}
            \IF{$u = p$}
                \STATE \textbf{continue}
            \ELSIF{$u \notin \mathrm{visited}$}
                \STATE $\mathrm{visited}[u] \leftarrow t$; $t \leftarrow t+1$; $\mathrm{parent}(u) \leftarrow v$; push $(u, v)$
            \ELSIF{$\mathrm{visited}[u] < \mathrm{visited}[v]$}
                \STATE $P \leftarrow$ vertices from $v$ up to $u$ along parent pointers
                \STATE append $(v, u, P)$ to $B$ \COMMENT{$P$ is the cycle closed by $(u,v)$}
            \ENDIF
        \ELSE
            \STATE pop
        \ENDIF
    \ENDWHILE
\ENDFOR
\RETURN $B$
\end{algorithmic}
\end{algorithm}

\begin{figure}[htbp]
\centering
\begin{tikzpicture}[x=1.05cm,y=1.05cm,
    vtx/.style={circle,draw,minimum size=6mm,inner sep=0pt,font=\footnotesize},
    onc/.style={vtx,fill=gray!30},
    kill/.style={vtx,fill=gray!55,very thick},
    gone/.style={circle,draw=gray!45,densely dotted,minimum size=6mm,inner sep=0pt,
                 font=\footnotesize,text=gray!55},
    disc/.style={font=\tiny,gray!70,inner sep=1pt},
    lbl/.style={font=\scriptsize,align=center,text width=4.35cm},
    >={Stealth[width=4pt,length=4pt]}]

% ================= (a) the traversal =================
\begin{scope}
  \node[vtx] (r) at (0,2.5)     {$r$};
  \node[vtx] (u) at (-0.95,1.6) {$u$};
  \node[vtx] (w) at (-0.95,0.7) {$w$};
  \node[vtx] (v) at (-0.95,-0.2){$v$};
  \node[vtx] (x) at (0.95,1.6)  {$x$};
  \node[vtx] (y) at (0.95,0.7)  {$y$};
  \node[vtx] (z) at (0.95,-0.2) {$z$};
  \node[disc,left=1pt of r]  {0}; \node[disc,left=1pt of u] {1};
  \node[disc,left=1pt of w]  {2}; \node[disc,left=1pt of v] {3};
  \node[disc,right=1pt of x] {4}; \node[disc,right=1pt of y]{5};
  \node[disc,right=1pt of z] {6};
  \draw[->] (r)--(u); \draw[->] (u)--(w); \draw[->] (w)--(v);
  \draw[->] (r)--(x); \draw[->] (x)--(y); \draw[->] (y)--(z);
  \draw[densely dashed,thick] (v) to[bend right=60] (u);
  \draw[densely dashed,thick] (z) to[bend left=60]  (x);
  \node[lbl,anchor=north] at (0,-0.9) {(a) one depth-first search. Solid: tree edges, numbered by discovery time. Dashed: the two back edges, each running to an ancestor ($1<3$ and $4<6$).};
\end{scope}

% ================= (b) what the back edges certify =================
\begin{scope}[xshift=4.9cm]
  \node[vtx] (r2) at (0,2.5)     {$r$};
  \node[onc] (u2) at (-0.95,1.6) {$u$};
  \node[onc] (w2) at (-0.95,0.7) {$w$};
  \node[onc] (v2) at (-0.95,-0.2){$v$};
  \node[onc] (x2) at (0.95,1.6)  {$x$};
  \node[onc] (y2) at (0.95,0.7)  {$y$};
  \node[onc] (z2) at (0.95,-0.2) {$z$};
  \draw[gray!45] (r2)--(u2) (r2)--(x2);
  \draw[very thick] (u2)--(w2) (w2)--(v2) (x2)--(y2) (y2)--(z2);
  \draw[densely dashed,very thick] (v2) to[bend right=60] (u2);
  \draw[densely dashed,very thick] (z2) to[bend left=60]  (x2);
  \node[lbl,anchor=north] at (0,-0.9) {(b) each back edge certifies exactly one cycle: $u$--$w$--$v$ and $x$--$y$--$z$. The search returns $\mu = m-n+c$ of them, not one per cycle of the graph.};
\end{scope}

% ================= (c) select, delete, re-prune =================
\begin{scope}[xshift=9.8cm]
  \node[vtx]  (r3) at (0,2.5)     {$r$};
  \node[kill] (u3) at (-0.95,1.6) {$u$};
  \node[gone] (w3) at (-0.95,0.7) {$w$};
  \node[gone] (v3) at (-0.95,-0.2){$v$};
  \node[onc]  (x3) at (0.95,1.6)  {$x$};
  \node[onc]  (y3) at (0.95,0.7)  {$y$};
  \node[onc]  (z3) at (0.95,-0.2) {$z$};
  \draw[gray!45] (r3)--(u3) (r3)--(x3);
  \draw[gray!45] (u3)--(w3) (w3)--(v3);
  \draw[gray!45,densely dashed] (v3) to[bend right=60] (u3);
  \draw[very thick] (x3)--(y3) (y3)--(z3);
  \draw[densely dashed,very thick] (z3) to[bend left=60] (x3);
  \node[lbl,anchor=north] at (0,-0.9) {(c) the rule picks one vertex of one certified cycle ($u$ here) and deletes it. Its cycle is gone; $w$ and $v$ leave with the 2-core re-prune. The other cycle survives, and the loop repeats on it.};
\end{scope}

\end{tikzpicture}
\caption[The DFS back-edge decycler]{One iteration of the Phase-1 procedure, on
a graph with $n=7$, $m=8$ and hence $\mu = 2$. A depth-first search of the
2-core (a) numbers each vertex by discovery time and reports a \emph{back edge}
whenever it meets an already-visited vertex with a smaller number, that is, an
ancestor. Each back edge certifies exactly one cycle (b), and there are
$\mu = m-n+c$ of them rather than one per cycle of the graph, which is why the
search costs $O(n+m)$ however many cycles the graph contains. The selection
rules of Sections~\ref{sec:baseline-dfs}, \ref{sec:option1} and~\ref{sec:option2}
differ only in which certified cycle is processed and which of its vertices is
deleted (c); the 2-core is then rebuilt and the search repeated. Deleting one
vertex per iteration is what creates the per-cycle choice the first realisation
never made.}
\label{fig:back-edge}
\end{figure}
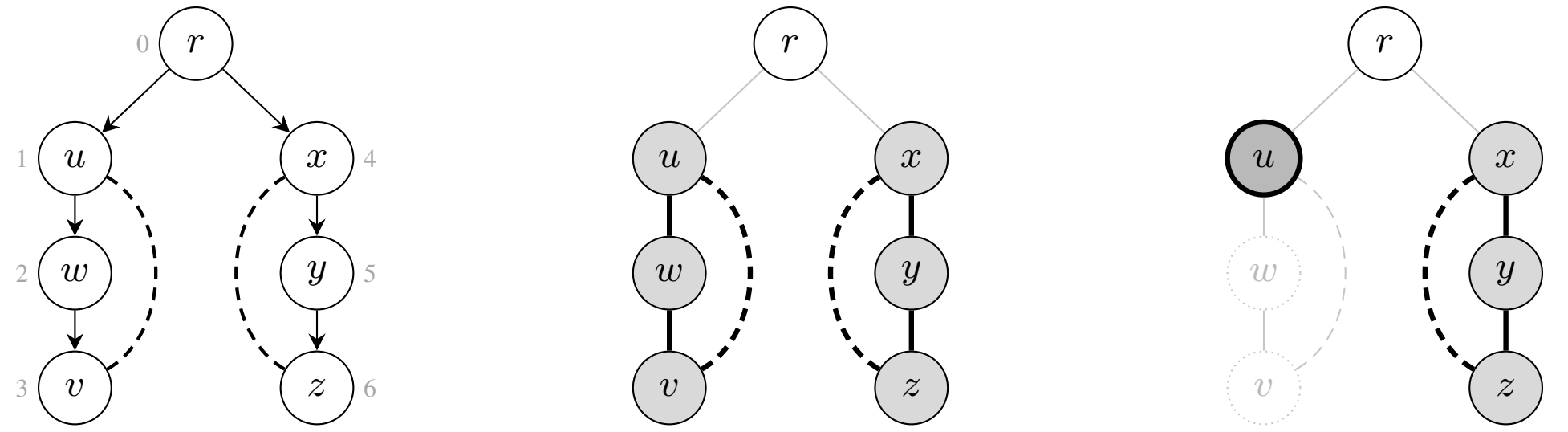

The traversal is iterative rather than recursive: on networks of even moderate
size the recursion depth of a recursive depth-first search exceeds typical
runtime stack limits. Two properties of the procedure deserve note because they
affect reproducibility. First, the rankings used by the two Collective-Influence
strategies defined below are recomputed from scratch at every iteration, on the
residual graph, rather than fixed once and consumed in batches. Second, the
outer loop visits the roots of the search, the vertices of the 2-core written $C$
in Algorithm~\ref{alg:back-edge}, in ascending label order, and explores each
vertex's neighbours in adjacency-list order, so which back edge is found first,
and hence the outcome of any tie among equally scored candidates, is fixed by the
labelling and by the order in which the adjacency lists were built. A removal
sequence is therefore a function of the labelled graph, and a re-run reproduces
it exactly; it is not a function of the unlabelled structure, since relabelling
the vertices can change which of several equally scored candidates is removed.

The cost of a pass is easy to overstate. A depth-first search does not enumerate
the cycles of the 2-core, of which there may be exponentially many; it reports
one back edge per \emph{independent} cycle, and by
Definition~\ref{def:cycle-rank} there are exactly $\mu = m - n + c$ of those on
the searched subgraph. On the Erd\H{o}s--R\'enyi 2-cores at seeds $42$, $43$ and
$44$ the search returns $255$, $263$ and $271$ back edges against 2-cores of
$379$, $371$ and $372$ vertices respectively. Extracting the 2-core and running
the search are each $O(n + m)$, so one pass of Algorithm~\ref{alg:back-edge}
costs $O(n + m)$ however many cycles the graph contains.

The expense lies in the outer loop rather than the search. Each pass deletes a
single vertex and then rebuilds the 2-core and the back-edge set from scratch, so
decycling a graph costs $O(|F|(n + m))$ for a returned set $F$. Since $|F|$ runs
between $14.3\%$ and $16.1\%$ of the vertices on the sparse synthetic families
and between $42.5\%$ and $61.4\%$ on the two real networks, that is $O(nm)$ in
the regime evaluated here. No incremental update is attempted, and
Section~\ref{sec:cost} explains why none is available.

Three selection strategies are benchmarked. The \emph{baseline} DFS rule and two
Collective-Influence-guided refinements, \emph{Max-Endpoint CI} and
\emph{Cycle-Path CI}, with a length-normalised variant of the latter,
\emph{Mean-Path CI}, included as a control. The refinements draw their scores
from Collective Influence at radius $1$,
\[
\mathrm{CI}_1(i) \;=\; (k_i - 1) \sum_{j \in \partial B(i,1)} (k_j - 1),
\]
the $\ell = 1$ case of the measure of \citet{morone2015}, where $k_i$ denotes
the degree of vertex $i$ and $\partial B(i,1)$ is simply its neighbourhood.
Degree- and coreness-based selection were also considered and are not among the
benchmarked strategies; they are not left unmeasured, but evaluated as controls
in Section~\ref{sec:ordering-not-score}, where the first isolates what the
influence score contributes over a score that is free.

\subsubsection{Baseline: Higher-Degree Endpoint of the First Back Edge}
\label{sec:baseline-dfs}

The baseline rule processes the first back edge $(u,v)$ the traversal encounters
and breaks the certified cycle at its higher-degree endpoint,
$\arg\max\{d(u), d(v)\}$. This is the minimal importance-aware refinement of the
first realisation's rule: it consults degree, which costs nothing to obtain,
since residual degrees are already maintained by the 2-core pruning of
Section~\ref{sec:2core}, so the rule costs $O(1)$ beyond the traversal itself. It
directly encodes the first lesson of Section~\ref{sec:requirements}: never spend
a removal on a degree-two vertex when a hub is available on the same cycle. Its
blind spots are equally clear. \emph{Which} cycle is processed is still dictated
by traversal order, and the candidate set is only the two endpoints of the back
edge, not the whole cycle.

\subsubsection{Max-Endpoint CI}
\label{sec:option1}

Max-Endpoint CI ranks the back edges discovered by the traversal by
$\max\{\mathrm{CI}_1(u), \mathrm{CI}_1(v)\}$, processes them in decreasing order
of this score, and breaks each processed cycle at the endpoint attaining the
maximum. The refinement over the baseline is twofold. First, the score considers
not just a vertex's own degree but the degrees of its neighbours:
$\mathrm{CI}_1$ multiplies a vertex's excess degree by the summed excess degree
of its neighbours, so it distinguishes a hub embedded among other hubs from a hub
whose neighbours are leaves, since only the former sustains connectivity beyond
its immediate neighbourhood. Second, ranking the back edges globally removes the
traversal-order arbitrariness in \emph{which} cycle is processed, so importance
now governs both free choices rather than only the second. Each score evaluation
costs $O(d(i))$ time. Section~\ref{sec:ordering-not-score} separates these two
refinements experimentally, and finds that the second carries the gain.

\subsubsection{Cycle-Path CI}
\label{sec:option2}

Cycle-Path CI scores each back edge by the total influence of the cycle it
certifies: for a back edge $(u,v)$ whose cycle consists of the tree path $P$ from
$v$ to $u$ closed by the edge, the score is
\[
\sum_{i \in P} \mathrm{CI}_1(i),
\]
back edges are processed in decreasing order of this score, and the top-ranked
cycle is broken at its maximum-$\mathrm{CI}_1$ vertex. Whereas Max-Endpoint CI
judges a cycle by its best single vertex, Cycle-Path CI judges the cycle as a
whole: a cycle threading entirely through an influential region outranks a cycle
that merely touches one hub, so removals are spent on the cycles that are
themselves most structurally central. The price is that scoring a back edge costs
time proportional to the length of its tree path times the average degree along
it, rather than the constant number of endpoint evaluations Max-Endpoint CI
needs. Section~\ref{sec:length-confound} shows that in practice this score is
dominated by path length rather than by influence.

\subsection{Phase 2: Greedy Collective-Influence Removal}
\label{sec:phase2}

\subsubsection{Collective Influence on Trees}
\label{sec:ci-trees}

The Collective Influence of a vertex at radius $\ell$ is
\[
\mathrm{CI}_\ell(v) \;=\; \bigl(d(v)-1\bigr) \sum_{u \in \partial B(v,\ell)} \bigl(d(u)-1\bigr),
\]
where $\partial B(v,\ell)$ is the set of vertices at distance exactly $\ell$ from
$v$. The measure was derived by \citet{morone2015} from the optimal-percolation
formulation of dismantling: the influence of a vertex is obtained from the
stability analysis of a message-passing system on the network, and the derivation
is exact precisely when the ball $B(v,\ell)$ around each vertex is a tree. On
general graphs this locally-tree-like assumption fails wherever short cycles
exist, and $\mathrm{CI}_\ell$ is only an approximation whose quality degrades
with the density of loops.

This bears on the second phase of the pipeline. Phase~1 terminates when the 2-core of the surviving
graph is acyclic. By Lemma~\ref{lem:cycles-in-2core} every cycle lies within the
2-core, so at that moment the surviving graph is a forest: this is the handoff.
Deleting a vertex from a forest leaves a forest, and Phase~2 only ever deletes
vertices; the graph on which the measure is evaluated is therefore a forest not
merely at the handoff but at every subsequent step. So within a decycling-based
strategy the assumption behind Collective Influence is not approximated but met
outright, for the whole of Phase~2: there are no cycles left for it to fail on.

Two conditions fix how far this reaches. Exactness is a property of the
\emph{score} rather than of the greedy policy built on it, so removing the
$\mathrm{CI}$-maximal vertex at each step is not thereby an optimal
fragmentation policy. And it holds only for strategies that decycle first. The
greedy Collective Influence baseline of Section~\ref{sec:setup} runs the
identical rule on the raw input graph, which is not a forest, so there the
measure is an approximation like any other.

The radius is the weaker point. The implemented rule uses $\ell = 1$, where the
tree assumption asks very little: $\mathrm{CI}_1$ inspects only a vertex and its
immediate neighbours, so all it requires is that no two neighbours of a vertex
are adjacent. Exactness at $\ell = 1$ is real but weak, and it is not the
principal justification for the design. The same caution applies inside Phase~1,
where the $\mathrm{CI}_1$ rankings of Section~\ref{sec:selection} are computed on
the still-cyclic 2-core and Collective Influence is a heuristic and nothing more.

The measure also responds to the quantity the dismantling objective is scored on,
namely component size. Every leaf has $d(v) - 1 = 0$ and hence
$\mathrm{CI}_\ell(v) = 0$, so CI concentrates on internal vertices, weighted by
the branching mass at distance $\ell$, a sensitivity to component \emph{mass}
that the eccentricity-based center criterion lacks. A tree consisting of a long
path with a star attached at one end illustrates the difference: the center of
that tree sits midway along the path and its removal barely shrinks the largest
component, whereas the star's hub dominates every path vertex in Collective
Influence, and its removal dissolves the star in a single step.

\subsubsection{The Greedy Collective-Influence Fill}
\label{sec:ci-fill}

Phase~2 is the greedy Collective-Influence rule of Algorithm~\ref{alg:ci-fill}.
Its behaviour differs from what the term ``fragmentation'' might suggest, and the
differences are deliberate.

It is \emph{global rather than per-component}. The rule does not identify an
oversized component and work within it; at each step it scores every surviving
vertex of the whole residual graph and deletes the global maximum. It uses
\emph{radius $1$ only}: the score is
\[
\mathrm{CI}_1(v) \;=\; \bigl(d(v)-1\bigr) \sum_{u \in N(v)} \bigl(d(u)-1\bigr),
\]
evaluated on the graph of surviving vertices, so no radius parameter $\ell$ is
exposed and the boundary $\partial B(v,\ell)$ of the general definition collapses
to the neighbourhood $N(v)$. And it carries \emph{no threshold}: the loop does
not stop when components fall below a size $\tau$ but continues until no vertex
remains. Running to exhaustion serves the measurement, not the algorithm: it is
done so that the strategies can be compared, not because dismantling requires it.
It yields a removal sequence of full length $N$, which is what makes the
largest-component curve of a decycling-based strategy comparable with those of
the single-rule baselines over the entire range $f \in [0,1]$; a sequence that
halted at a threshold would be defined only on a prefix of that range.

Ties are resolved by vertex label: among the vertices attaining the maximum
score, the one with the smallest label is deleted. No degree or other structural
criterion is consulted. The rule is not a formality, because ties are common:
every leaf and every isolated vertex scores exactly zero, since each carries an
excess degree $d(v) - 1 = 0$ or has no neighbours to sum over. Late in the fill,
when the residual consists largely of such vertices, nearly every step is a tie,
and the order in which they are removed is settled by the labelling rather than
by anything structural. Fixing the rule this way makes the removal sequence a
function of the labelled graph, so a run reproduces exactly; it does not make the
late-stage ordering structurally meaningful, and nothing in what follows treats it
as though it were.

\begin{algorithm}[htbp]
\caption{Greedy radius-1 Collective-Influence fill (Phase 2 as implemented)}
\label{alg:ci-fill}
\begin{algorithmic}[1]
\REQUIRE Residual graph $H$ (the Phase-1 output), removal list $R$ built so far
\ENSURE $R$ extended to a removal sequence covering every vertex of $H$
\WHILE{$H$ has at least one vertex}
    \FOR{each $v \in V(H)$}
        \STATE $\mathrm{CI}_1(v) \leftarrow \bigl(d_H(v)-1\bigr) \sum_{u \in N_H(v)} \bigl(d_H(u)-1\bigr)$
    \ENDFOR
    \STATE $v^{*} \leftarrow \arg\max_{v \in V(H)} \mathrm{CI}_1(v)$ \COMMENT{ties broken by smallest vertex label}
    \STATE append $v^{*}$ to $R$; delete $v^{*}$ from $H$
\ENDWHILE
\RETURN $R$
\end{algorithmic}
\end{algorithm}

Each iteration deletes one vertex from a finite graph, so the procedure
terminates after exactly $|V(H)|$ iterations and returns a sequence covering
every vertex.

That last point governs how the experimental results of
Section~\ref{sec:experiments} must be read. Algorithm~\ref{alg:ci-fill} is not a
component of one strategy among several: it is the identical continuation of
\emph{every} strategy that runs a decycling phase, invoked at the moment the
graph becomes acyclic. The three selection rules of Section~\ref{sec:selection}
differ from one another only before that moment. The greedy Collective Influence
baseline is this same procedure applied from the outset, with no decycling phase
preceding it. How much of each sequence is consequently attributable to the
strategy under study, rather than to this shared rule, is measured in
Section~\ref{sec:shared-tail}.

\subsection{Computational Cost}
\label{sec:cost}

The two phases have very different cost profiles, and the difference is worth
setting out because it is not the one the complexity bounds suggest at first
reading.

Phase~2 admits an exact optimisation. A vertex's score
$\mathrm{CI}_1(u) = (d(u)-1)\sum_{w \in N(u)}(d(w)-1)$ depends on its own degree
and on the degrees of its neighbours, and on nothing else, so deleting $v$ can
change the score only of vertices within distance two of $v$. A vertex three hops
away has the same degree and the same neighbours with the same degrees, and its
score is unchanged. Rescoring that ball and keeping the remaining candidates in a
heap therefore computes the same sequence as a full rescan while touching a small
fraction of the graph. A full rescan costs $O(m)$ per removal and $O(Nm)$ over
the sequence; the incremental form costs $O(|B_2(v^{*})| \log N)$ per removal,
which is only $O(Nm)$ in the worst case, when some ball spans the graph, and far
below it on the sparse residuals that arise here. Table~\ref{tab:cost} reports
the effect.

\begin{table}[htbp]
\centering
\caption{Cost of the greedy fill, full rescan against incremental update. Both
produce identical removal sequences; the rescan figures are the cost of the
implementation as originally written. Erd\H{o}s--R\'enyi graphs at
$\langle k \rangle = 3$, single instances, wall clock on one machine.}
\label{tab:cost}
\small
\begin{tabular}{lrrr}
\toprule
Graph & full rescan & incremental & speedup \\
\midrule
$n = 500$            & $0.50$\,s  & $0.005$\,s & $107\times$ \\
$n = 1000$           & $2.08$\,s  & $0.01$\,s  & $219\times$ \\
$n = 2000$           & $8.27$\,s  & $0.02$\,s  & $435\times$ \\
$n = 4000$           & $32.83$\,s & $0.04$\,s  & $870\times$ \\
Add Health           & $30.3$\,s  & $0.15$\,s  & $202\times$ \\
CollegeMsg proxy     & $5.6$\,s   & $0.23$\,s  & $25\times$ \\
\bottomrule
\end{tabular}
\end{table}

The equivalence is exact rather than approximate, and it has to be, since every
result in this paper was measured with the original rule. The tie-break is where
that could fail: every leaf and every isolated vertex scores zero, so late in a
fill almost every step is a tie, and the original resolved those to the smallest
vertex label only as a side effect of scanning a label-sorted list. A heap keyed
on $(-\mathrm{CI}_1, \text{label})$ reproduces that order, and the two
implementations were checked against one another on $145$ graphs, including
deliberately tie-heavy inputs, with no differences.

\textbf{Phase~1 admits no equivalent, and the reason is structural.} Its rule
ranks the back edges returned by a depth-first search of the current 2-core. A
depth-first search is a global traversal, and \emph{which} back edges it returns
depends on the path it walks, so the search must be repeated after every removal;
there is no local update because the quantity being computed is not local.
Substituting a materialised 2-core for the filtered subgraph the implementation
uses was tested and rejected: it altered the removal sequence on seven of twenty
graphs, because neighbour iteration order changed and the traversal consequently
found different back edges. Phase~1 therefore remains $O(|F|(n+m))$, and that
bound belongs to the rule as defined in Section~\ref{sec:selection} rather than
to any deficiency of the code.

The consequence for the pipeline is a reversal of the naive reading. Judged by
worst-case complexity the fill looks dominant, at $O(Nm)$ against $O(|F|(n+m))$.
Measured, it accounts for under one per cent of the running time and Phase~1 for
essentially all of it. Against CoreHD the pipeline costs $2.5$ times as much at
$n = 500$, rising to $3.3$ at $n = 2000$ and $8.6$ on Add Health, all of that
difference attributable to the decycling phase. Wall-clock figures are of course
specific to one machine and to an implementation in an interpreted language; what
they establish is the relative weight of the two phases, not an absolute cost.

\section{What Is and Is Not Established}
\label{sec:theory}

This section states two results, neither of them a contribution of this work.
The first is the decycling bound carried by the local-ratio subroutine that
ships with the accompanying code; the second is the classical component-size
bound for centroid removal, which fixes what the greedy fill of
Section~\ref{sec:ci-fill} gives up. Each belongs to a specific subroutine, and
this section says which.

For the evaluated Phase~1, the DFS back-edge decycler of
Section~\ref{sec:selection} in each of its variants, what is established is only
that the set it removes is a valid feedback vertex set, that the procedure
terminates, and that it runs in polynomial time. Validity is immediate from the
termination test: the loop of Algorithm~\ref{alg:back-edge} exits only when the
2-core of the residual graph is acyclic, which by
Lemma~\ref{lem:cycles-in-2core} means the residual graph is a forest.
Termination is immediate because each iteration deletes one vertex from a finite
graph. The running time is the $O(|F|(n+m))$ of Section~\ref{sec:selection}.
Its performance rests on the empirical evidence of
Section~\ref{sec:experiments}.

\subsection{A Decycling Bound for the Local-Ratio Subroutine}
\label{sec:decycling-bound}

The codebase accompanying this work contains an implementation of the
local-ratio $2$-approximation for feedback vertex set,
\texttt{netdis.fvs\_local\_ratio.bbf\_fvs}, which initialises unit vertex
weights and always executes the reverse-delete minimality pass. For that
subroutine the classical guarantee holds.

\begin{theorem}[Decycling bound, subroutine only]
\label{thm:bbf-decycling-bound}
Let $G$ be a finite undirected graph and let $F$ be the vertex set returned by
\texttt{bbf\_fvs}$(G)$ under unit vertex weights. Then $F$ is a feedback vertex
set of $G$ and
\[
|F| \;\le\; 2\,|F^{*}|,
\]
where $F^{*}$ is a minimum-cardinality feedback vertex set of $G$.
\end{theorem}

The theorem is not a result of this work and is not proved here. It is the
guarantee of~\citet{bafna1999} for the algorithm they call FEEDBACK, specialised
to unit vertex weights. Their argument turns on the observation, their
Lemma~4.2, that under a degree-proportional weight function the weight of
\emph{any} minimal feedback vertex set is within a factor of two of the optimum;
\citet{chudak1998} later recovered the same factor for this algorithm and for
that of~\citet{becker1996} from a single primal--dual formulation.

The statement carries four qualifications. The bound is on the
\emph{cardinality} of the returned set and on nothing else; it holds for
\emph{unit weights}, as this subroutine initialises them; it requires the
\emph{reverse-delete minimality pass}, minimality being a hypothesis of the
underlying analysis rather than a by-product of the weight loop; and it applies
to the \emph{decycling phase alone}, the removal of a set that destroys every
cycle.

\medskip
\noindent\textbf{This bound does not extend to any pipeline in this paper.} A
feedback vertex set of at most twice the optimum does not yield a dismantling
sequence of any bounded quality. The dismantling objective concerns the size of
the largest surviving component as vertices are removed in order, and
Theorem~\ref{thm:bbf-decycling-bound} constrains neither the collapse fraction
$f^{*}$, nor the area under the largest-component curve, nor the number of
removals required to bring every component below a threshold. Nor do the
per-phase bounds compose: the optimal dismantling set is not the union of an
optimal feedback vertex set and an optimal forest fragmentation, and the
fragmentation instance handed to Phase~2 is itself an artifact of what Phase~1
chose to remove.

\medskip
\noindent\textbf{This subroutine is a comparator, not the evaluated method.}
\texttt{bbf\_fvs} appears as a row in Tables~\ref{tab:vs-first}
and~\ref{tab:fixed-completion}, scored under the same protocol as every other
decycler so that the two can be read against one another. It is not one of the
proposed strategies, and no result reported for them depends on it.
What Theorem~\ref{thm:bbf-decycling-bound} governs travels no further than that
row: the cardinality of the set \texttt{bbf\_fvs} returns.

The comparison belongs here, because it settles what the bound is and is not
good for. Run as a dismantler, with its set emitted first and then the greedy
fill of Algorithm~\ref{alg:ci-fill} that every decycling strategy hands off to,
\texttt{bbf\_fvs} reaches $f^{*} = 0.197$ on Erd\H{o}s--R\'enyi and $0.192$ on
the configuration model (Table~\ref{tab:vs-first}), while removing a decycling
set of $14.8\%$ and $14.5\%$ of the vertices, against $16.1\%$ for Baseline DFS
on both families. It is not the smallest in the comparison: Max-Endpoint CI
reaches acyclicity at $14.3\%$ to $14.5\%$. Those are competitive figures, ahead
of Baseline DFS on both families and behind Max-Endpoint CI on both. The
commonly assumed trade-off, that a worst-case bound is paid for in practice, is
therefore not visible on these instances. Neither, however, does the bounded
subroutine produce the result, and where the collapse happens shows why. The
largest component still holds $67.7\%$ and $68.6\%$ of the vertices at BBF's
acyclicity point, so the $5\%$ threshold is crossed well inside the shared fill
rather than during the removal of $F$. What the theorem governs is the size of
$F$; what determines $f^{*}$ is a phase the theorem says nothing about. This is
the shared-tail effect of Section~\ref{sec:shared-tail}, arriving from the one
method in this comparison that carries a proof.

One consequence should be stated precisely, because it is what the decycler of
Section~\ref{sec:selection} is built to avoid rather than any deficit in
performance. A feedback vertex set fixes which vertices leave and not the order
in which they go, so scoring BBF as a dismantler at all requires an ordering the
algorithm does not supply; ascending label order is used here. On these
instances that choice provably cannot move $f^{*}$, since the largest component
exceeds the collapse threshold throughout the removal of $F$, so the crossing
must fall inside the order-independent fill, and twenty random permutations of
$F$ per instance leave $f^{*}$ unchanged to three decimals. It does move the
area under the curve, across $[0.138, 0.140]$ on Erd\H{o}s--R\'enyi and
$[0.135, 0.138]$ on the configuration model.
Theorem~\ref{thm:bbf-decycling-bound} is therefore not an unused result here,
but its role is narrower than a dismantling guarantee: it bounds $|F|$ and
nothing downstream of it, leaving an arbitrary degree of freedom in the reported
curve. That is the argument for a decycler emitting an ordering rather than a
set, not that the set dismantles badly, which on this evidence it does not.

\subsection{A Component-Size Bound for Centroid Removal}
\label{sec:centroid}

Phase~2 could have kept a tree-structural criterion instead of an importance
score, and the comparison is worth stating because it is a genuine trade-off
rather than a strict improvement. The tree \emph{center} carries no bound on
component sizes at all: on a long path with a star attached at one end, the
center sits midway along the path and its removal barely shrinks the largest
component. The \emph{centroid} does carry one, by the following classical result
of \citet{jordan1869}.

\begin{theorem}[Centroid bound]
\label{thm:centroid-bound}
Every tree $T$ on $n_T$ vertices contains a vertex $c$, a centroid, such that
every connected component of $T - c$ has at most $\lfloor n_T/2 \rfloor$
vertices.
\end{theorem}

\noindent\emph{Proof.} For a vertex $v$ of $T$, let $f(v)$ denote the number of
vertices in a largest component of $T - v$, and let $c$ minimize $f$. Suppose,
for contradiction, that $f(c) > \lfloor n_T/2 \rfloor$, i.e.\ $f(c) > n_T/2$.
Since the components of $T - c$ partition $n_T - 1$ vertices, at most one of them
can exceed $n_T/2$; let $C$ be that component, with $|C| = f(c)$, and let $u$ be
the unique neighbour of $c$ in $C$ (uniqueness holds because a second neighbour
would close a cycle). Consider $T - u$. Its components are of two kinds: (a) the
components of $C - u$, each of size at most $|C| - 1 = f(c) - 1$; and (b) the
component containing $c$, which consists of $c$ together with all components of
$T - c$ other than $C$, of size $n_T - |C| = n_T - f(c) < n_T/2 < f(c)$. Hence
$f(u) \le \max\{f(c) - 1,\; n_T - f(c)\} < f(c)$, contradicting the minimality of
$c$. Therefore $f(c) \le \lfloor n_T/2 \rfloor$. \hfill$\square$

The bound converts directly into a worst-case cost for a fragmentation phase
built on it.

\begin{lemma}[Cost of centroid recursion]
\label{lem:centroid-recursion}
Let $\tau \ge 1$. Recursive centroid removal fragments any tree on $n_T$
vertices into components of size at most $\tau$ using recursion depth at most
$\lceil \log_2 (n_T/\tau) \rceil$ and at most $2 n_T/(\tau + 1) - 1$ removals.
\end{lemma}

\noindent\emph{Proof.} By Theorem~\ref{thm:centroid-bound}, every component at
recursion depth $k$ has at most $n_T / 2^{k}$ vertices, which is at most $\tau$
once $k \ge \log_2(n_T/\tau)$; this gives the depth bound. For the removal count,
let $r(s)$ be the number of removals the recursion performs on a tree of $s$
vertices; we show $r(s) \le 2s/(\tau+1) - 1$ for all $s > \tau$ by strong
induction (for $s \le \tau$, $r(s) = 0$). Removing the centroid of a tree of size
$s > \tau$ costs one removal and leaves components $s_1, \dots, s_p$ with
$\sum_i s_i = s - 1$ and each $s_i \le \lfloor s/2 \rfloor$; let $k$ be the
number of components with $s_i > \tau$, the only ones that recurse. If $k = 0$,
then $r(s) = 1 \le 2s/(\tau+1) - 1$, since $s \ge \tau + 1$. If $k = 1$, with the
recursing component of size $s_1 \le \lfloor s/2 \rfloor$, the inductive
hypothesis gives
\[
r(s) \;\le\; 1 + \frac{2 s_1}{\tau+1} - 1 \;\le\; \frac{s}{\tau+1} \;\le\; \frac{2s}{\tau+1} - 1,
\]
the last step again using $s \ge \tau + 1$. If $k \ge 2$, then since
$\sum_i s_i \le s - 1$,
\[
r(s) \;\le\; 1 + \sum_{i \,:\, s_i > \tau} \left( \frac{2 s_i}{\tau+1} - 1 \right)
\;\le\; 1 + \frac{2(s-1)}{\tau+1} - k
\;\le\; \frac{2s}{\tau+1} - 1 - \frac{2}{\tau+1}
\;<\; \frac{2s}{\tau+1} - 1.
\]
In all cases $r(s) \le 2s/(\tau+1) - 1$, and the lemma follows with
$s = n_T$. \hfill$\square$

The greedy rule of Algorithm~\ref{alg:ci-fill} does not inherit this guarantee,
because a $\mathrm{CI}_1$-maximal vertex need not split its component in a
balanced way. Phase~2 thus forces a choice between the rule with the worst-case
bound (the centroid) and the rule with the exact importance measure on trees
(Collective Influence), and which one wins in practice is settled empirically in
Section~\ref{sec:vs-first}.

\section{Experimental Evaluation}
\label{sec:experiments}

\subsection{Setup}
\label{sec:setup}

\paragraph{Synthetic graphs.} Two random-graph families are used.
Erd\H{o}s--R\'enyi graphs $G(n,p)$ are generated with
$p = \langle k \rangle/(n-1)$, and configuration-model graphs are generated from
a Poisson$(\langle k \rangle)$ degree sequence with parallel edges merged and
self-loops discarded, which places the realised average degree slightly below the
nominal one. Unless a sweep says otherwise, synthetic results use $n = 500$ and
$\langle k \rangle = 3.0$, averaged over twenty independently seeded instances
(seeds $42$--$61$). Section~\ref{sec:shared-tail} sweeps $n$ from $500$ to
$4000$ and Section~\ref{sec:density} sweeps $\langle k \rangle$ from $2.5$ to
$6.0$, each at twenty seeded instances per point.

\paragraph{Real networks.} Two real networks are used. \emph{Add Health} is the
Konect \texttt{moreno\_health} network, derived from the National Longitudinal
Study of Adolescent Health friendship survey; after reduction to its largest
connected component it has $2539$ vertices and $10{,}455$ edges.
\emph{CollegeMsg proxy} is the SNAP CollegeMsg private-message network trimmed to
its $500$ highest-coreness vertices. It is used in place of the MIT Reality
Mining proximity network, which is not openly redistributable, and it is
substantially denser than that network; results reported under this name
characterise the proxy and should not be read as results for Reality Mining
itself. Both networks are considerably denser than the synthetic operating point,
and both are single instances, so they carry no instance spread.

Table~\ref{tab:networks} gives the structural statistics. It includes the cycle
rank ratio $\mu/n$ deliberately: by Lemma~\ref{lem:fvs-density-bound} this single
column predicts where feedback-vertex-set-based dismantling must be expensive,
and the density gap between the synthetic and real instances is visible in it
before any result curve is examined.

\begin{table}[htbp]
\centering
\caption{Structural statistics of the evaluated networks. Synthetic rows are
means over twenty instances (seeds 42--61); real-network rows are single
instances after reduction to the largest connected component.}
\label{tab:networks}
\small
\begin{tabular}{lrrrrrr}
\toprule
Network & $n$ & $m$ & $\langle k \rangle$ & $\mu$ & 2-core & $\mu/n$ \\
\midrule
Erd\H{o}s--R\'enyi ($\langle k \rangle = 3$) & 500 & 747.0 & 2.99 & 275.2 & 384.0 & 0.550 \\
Configuration model ($\langle k \rangle = 3$) & 500 & 744.0 & 2.98 & 271.1 & 379.1 & 0.542 \\
Add Health & 2539 & 10455 & 8.24 & 7917 & 2471 & 3.118 \\
CollegeMsg proxy & 500 & 7967 & 31.87 & 7468 & 500 & 14.936 \\
\bottomrule
\end{tabular}
\end{table}

\paragraph{Baselines.} Four baselines are used. \emph{Random} removes vertices in
uniformly random order. \emph{CoreHD}~\citep{zdeborova2016} iteratively removes
the highest-degree vertex of the current 2-core, with degree measured within the
2-core subgraph and the core recomputed after every removal. \emph{Collective
Influence} greedily removes the vertex maximizing $\mathrm{CI}_\ell$
\citep{morone2015}; the benchmarked configuration uses radius $\ell = 1$
throughout, recomputed after every removal, which makes it exactly
Algorithm~\ref{alg:ci-fill} applied from the outset with no decycling phase.
\emph{WEAK-NEIGHBOR}~\citep{schmidt2019} is included as a further comparator,
since it, CoreHD and the DFS back-edge decycler are all sequence-based decyclers
and are therefore directly comparable at the acyclicity handoff. The local-ratio
subroutine \texttt{bbf\_fvs} of Section~\ref{sec:decycling-bound} appears as a
fifth comparator where a decycling set is what is being compared.

\paragraph{Ablation baseline.} The first realisation of
Section~\ref{sec:requirements} is carried forward as an ablation baseline: one
component is reverted while everything else is held fixed, so that whatever
changes in the results can be attributed to that component. Here the reverted
component is the Phase-1 procedure.

\paragraph{Metrics.} The collapse fraction $f^{*}$ and the area under the LCC
curve over $f \in [0,1]$, both defined in Section~\ref{sec:preliminaries}. Lower
is better for both.

\subsection{Comparison with the First Realisation}
\label{sec:vs-first}

Table~\ref{tab:vs-first} places the selection strategies of
Section~\ref{sec:selection} against the ablation baseline on both synthetic
families.

\begin{table}[htbp]
\centering
\caption{The proposed pipeline against the first realisation on sparse synthetic
graphs ($n = 500$, $\langle k \rangle = 3$, means over seeds 42--61). Lower is
better throughout. The handoff column is the fraction of vertices removed when
the graph first becomes acyclic. The BBF row is scored under the same protocol
as the others, its set first and then the shared greedy fill, since the algorithm
returns no ordering of its own; see Section~\ref{sec:decycling-bound}.}
\label{tab:vs-first}
\small
\begin{tabular}{lccccc}
\toprule
 & \multicolumn{2}{c}{Erd\H{o}s--R\'enyi} & \multicolumn{2}{c}{Configuration model} & \\
Method & $f^{*}$ & AUC & $f^{*}$ & AUC & handoff \\
\midrule
First realisation (ablation) & 0.306 & 0.168 & 0.289 & 0.166 & --- \\
BBF $2$-approximation & 0.197 & 0.139 & 0.192 & 0.137 & 14.5--14.8\% \\
Baseline DFS & 0.203 & 0.143 & 0.205 & 0.143 & 16.1\% \\
Max-Endpoint CI & \textbf{0.191} & \textbf{0.136} & 0.196 & 0.136 & 14.3--14.5\% \\
Cycle-Path CI & 0.192 & \textbf{0.136} & 0.196 & \textbf{0.134} & 14.3--14.4\% \\
Mean-Path CI & 0.196 & 0.139 & \textbf{0.195} & 0.137 & 14.3--14.6\% \\
\bottomrule
\end{tabular}
\end{table}

The improvement over the first realisation is substantial and it is almost
entirely attributable to the change of decycling procedure rather than to the
choice of importance score. Every variant of the new Phase~1 reaches acyclicity
after removing between $14.3\%$ and $16.1\%$ of the vertices, where the first
realisation's decycler committed approximately $72\%$
(Section~\ref{sec:requirements}); the collapse fraction falls correspondingly,
from $0.306$ to between $0.191$ and $0.203$ on Erd\H{o}s--R\'enyi graphs. The gap
between the crudest new variant (Baseline DFS, $0.203$) and the best
(Max-Endpoint CI, $0.191$) is $1.2$ percentage points, against the $10.3$-point
gap between Baseline DFS and the first realisation. Replacing a procedure that
never chose with one that chooses at all is thus worth roughly eight times as
much as refining the criterion by which it chooses.

Because both phases changed at once, this comparison also settles the trade-off
of Section~\ref{sec:centroid} in the only form the evidence supports: the
pipeline that replaced recursive center removal with the greedy Collective
Influence fill dismantles far better, and the rule carrying the centroid's
worst-case bound was not the rule that won. That said, an ablation separating the
Phase-1 change from the Phase-2 change was not performed. No
configuration exists in the evaluated code that pairs the new Phase~1 with the
old Phase~2 or vice versa. The comparison in Table~\ref{tab:vs-first} is
therefore between two complete pipelines, and the attribution offered in the
preceding paragraph rests on the handoff column, a Phase-1 quantity, not on a
controlled ablation. Section~\ref{sec:fixed-completion} supplies the controlled
comparison that the handoff column only gestures at.

\subsection{Comparison with Baselines}
\label{sec:vs-baselines}

\begin{figure}[htbp]
    \centering
    \includegraphics[width=\textwidth]{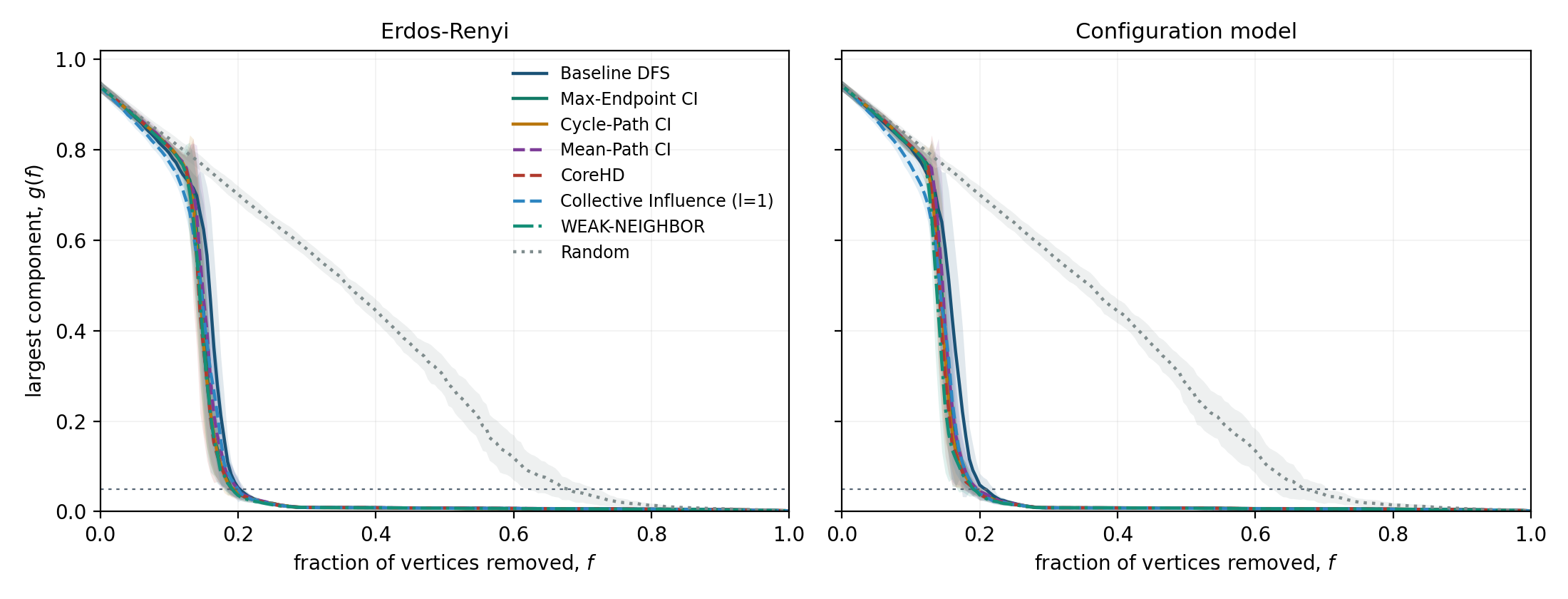}
    \caption{Dismantling curves on the two synthetic families ($n = 500$,
    $\langle k \rangle = 3$), for the selection strategies of
    Section~\ref{sec:selection} together with the baselines. Curves are means
    over twenty seeded instances (seeds 42--61) with one-standard-deviation
    bands. Two features are worth reading off directly. The seven structural
    methods are visually near-indistinguishable over the whole range, separating
    clearly only from random removal; and every handoff marker lies to the
    \emph{left} of its own curve's collapse, so each strategy has already
    surrendered control to the shared fill before its largest component breaks
    up. Both are the visual form of the shared-tail finding of
    Section~\ref{sec:shared-tail}.}
    \label{fig:synthetic-curves}
\end{figure}

Table~\ref{tab:baselines} reports the selection strategies against the external
baselines on all four networks, and Figure~\ref{fig:synthetic-curves} plots the
corresponding synthetic curves.

\begin{table}[htbp]
\centering
\caption{Collapse fraction $f^{*}$ against the baselines. Synthetic columns are
means over twenty seeded instances (seeds 42--61); real-network columns are
single instances. Lower is better.}
\label{tab:baselines}
\small
\begin{tabular}{lcccc}
\toprule
Method & ER & CFG & Add Health & CollegeMsg proxy \\
\midrule
Baseline DFS & 0.203 & 0.205 & 0.458 & 0.626 \\
Max-Endpoint CI & \textbf{0.191} & 0.196 & 0.444 & 0.602 \\
Cycle-Path CI & 0.192 & 0.196 & 0.443 & 0.590 \\
Mean-Path CI & 0.196 & 0.195 & 0.450 & 0.614 \\
CoreHD & 0.193 & \textbf{0.191} & 0.430 & 0.588 \\
Collective Influence ($\ell=1$) & 0.197 & \textbf{0.191} & 0.440 & 0.602 \\
WEAK-NEIGHBOR & \textbf{0.191} & \textbf{0.191} & \textbf{0.427} & \textbf{0.576} \\
Random & 0.672 & 0.665 & 0.829 & 0.932 \\
\bottomrule
\end{tabular}
\end{table}

The conclusion this table supports is that the proposed strategies are
competitive with, but do not beat, the established heuristics. Max-Endpoint CI
attains the joint-lowest collapse fraction on Erd\H{o}s--R\'enyi graphs, at
$0.191$, level with WEAK-NEIGHBOR and ahead of CoreHD's $0.193$; that margin is
$0.2$ percentage points over twenty instances and is well inside the
instance-to-instance spread, so it should not be read as a win. On the
configuration model CoreHD, Collective Influence and WEAK-NEIGHBOR all reach
$0.191$ against the best CI-DFS variant's $0.195$. On both real networks
WEAK-NEIGHBOR is best and CoreHD second, and no CI-DFS variant places first: the
best of them trails WEAK-NEIGHBOR by $1.6$ points on Add Health and $1.4$ points
on the CollegeMsg proxy. Across the four networks the proposed strategies place
first on none. Against random removal, all seven structural methods are separated
by a wide margin on every network.

The compression visible in Figure~\ref{fig:synthetic-curves} is itself the
finding. On the CollegeMsg proxy, which is not plotted here, the separation
between the selection strategies and the baselines is not visible at the scale of
the plot for the first half of the removal sequence.
Section~\ref{sec:density} explains why the field compresses as density rises.

\subsection{Isolating the Decycler under a Fixed Optimal Completion}
\label{sec:fixed-completion}

The two comparisons above share a defect, and it is the one
Section~\ref{sec:shared-tail} measures: every decycling strategy hands off to the
same greedy fill, on the synthetic families that fill is $84$ to $86\%$ of each
sequence, and the collapse point falls inside it in all fourteen
method-by-graph cells. Tables~\ref{tab:vs-first} and~\ref{tab:baselines}
therefore rank the residual each method left at handoff at least as much as they
rank the decisions it made. A margin of a few tenths of a point read off those
tables cannot be attributed to the selection rule with any confidence.

This section removes the confound rather than restating it. The idea is to hold
the second phase \emph{fixed and optimal} across every method, so that the only
quantity distinguishing two strategies is the forest each one leaves behind. Two
facts make this possible. Phase~1 terminates at acyclicity, so the graph handed
to Phase~2 is always a forest; and while dismantling a general graph is NP-hard
(Section~\ref{sec:preliminaries}), breaking a \emph{forest} into components of
size at most $\tau$ with the fewest deletions is polynomial.
\citet{braunstein2016} give the algorithm, the second stage of their three-stage
Min-Sum method, as a dynamic program over the tree, computing the minimum number
of deletions in time $O(n\tau^{2})$. Table~\ref{tab:fixed-completion} substitutes
that exact optimum for the greedy fill of Algorithm~\ref{alg:ci-fill},
identically for every strategy. The dynamic program is theirs and is not a
contribution of this work; what is new here is only its use as an experimental
control.

Because the optimum is a \emph{set} and carries no ordering, the quantity
tabulated is a removal count, $(|F| + \text{Phase-2 deletions})/n$, directly
comparable with a measured $f^{*}$ but supporting no area under the curve, the
same restriction that applies to the local-ratio subroutine in
Section~\ref{sec:decycling-bound}.

\begin{table}[htbp]
\centering
\caption{Every decycler completed by the same exact Phase~2, so the only
difference between rows is the forest each left at handoff ($n = 500$,
$\langle k \rangle = 3$, means over seeds 42--61, $\tau = 24$). Lower is better.
The Phase-2 column is the optimum of \citet{braunstein2016}'s tree-breaking
dynamic program.}
\label{tab:fixed-completion}
\small
\begin{tabular}{lcccccc}
\toprule
 & \multicolumn{3}{c}{Erd\H{o}s--R\'enyi} & \multicolumn{3}{c}{Configuration model} \\
Decycler & $|F|/n$ & Phase 2 & total & $|F|/n$ & Phase 2 & total \\
\midrule
Max-Endpoint CI     & 0.145 & 0.023 & 0.168 & 0.143 & 0.024 & 0.166 \\
WEAK-NEIGHBOR       & 0.142 & 0.023 & \textbf{0.165} & 0.139 & 0.023 & \textbf{0.162} \\
CoreHD              & 0.144 & 0.024 & 0.167 & 0.140 & 0.023 & 0.163 \\
BBF $2$-approximation & 0.148 & 0.022 & 0.170 & 0.145 & 0.023 & 0.168 \\
Baseline DFS        & 0.161 & 0.019 & 0.181 & 0.161 & 0.019 & 0.179 \\
\bottomrule
\end{tabular}
\end{table}

Three things follow, and they should be taken in order of how much they support.

First, and most strongly: \textbf{the fragmentation phase is not where
dismantling is decided.} The optimal Phase~2 costs between $0.019$ and $0.024$ of
the network whichever decycler precedes it, so between $86\%$ and $90\%$ of the
total is the feedback vertex set itself. The ranking by total is, in consequence,
exactly the ranking by $|F|$. This is a stronger statement than the shared-tail
finding of Section~\ref{sec:shared-tail}: it says that even with the fill made
optimal, and therefore with every trace of the shared-rule artifact removed, the
outcome is still governed almost entirely by the decycling step. Any future
effort spent refining forest fragmentation in this architecture is bounded above
by about two percentage points.

Second, under this comparison the CI-guided decycler is competitive but not
leading. On Erd\H{o}s--R\'enyi graphs Max-Endpoint CI totals $0.168$, against
WEAK-NEIGHBOR's $0.165$ and CoreHD's $0.167$, and ahead only of the local-ratio
decycler ($0.170$); on the configuration model it totals $0.166$ against $0.162$
and $0.163$. It is third of five on both families, within three tenths of a point
of the best. This is the cleanest comparison in the paper, since it is the only
one in which nothing but the decycling step varies, and what it shows is a
decycler that keeps pace with the strongest published ones without overtaking
them.

Third, the ordering of the decyclers is the ordering of their feedback vertex
sets, and on that ordering the CI-guided rule is not first. Its set is $0.145$
and $0.143$ of the vertices against WEAK-NEIGHBOR's $0.142$ and $0.139$ and
CoreHD's $0.144$ and $0.140$, and the totals follow in the same order. The claim
this section supports is therefore narrow: with the fragmentation phase held
optimal and identical, the procedure is within three tenths of a percentage
point of the best decycler in the comparison on both families, and behind it on
both. Section~\ref{sec:ordering-not-score} records that what separates the variants
at Phase~1 is the cycle ordering rather than the influence score.

One result is unambiguous across both families, and it is the one the redesign
was meant to obtain. Baseline DFS totals $0.181$ and $0.179$, between one and two
percentage points behind every other decycler, and its feedback vertex set is the
largest of the five. The gain from replacing the first realisation's decycler
with a scheme that removes one vertex per detected cycle is thus confirmed under
a protocol in which the shared fill cannot be responsible for it, while the
further gain from refining \emph{which} vertex that scheme selects remains, as
Section~\ref{sec:vs-first} reported, the smaller effect by a wide margin.

\section{Controls and Limits}
\label{sec:discussion}

Four findings qualify the results of the preceding section, and each of them
limits a claim that the design of the method might otherwise invite.

\subsection{Cycle-Path CI's Cycle Score Is a Length Score}
\label{sec:length-confound}

Cycle-Path CI ranks back edges by the unnormalized sum of $\mathrm{CI}_1$ along
the certified cycle path, which was intended to prefer cycles threading through
structurally central regions. Replaying the algorithm's own trajectory and
pooling every back edge it examined shows that the score is dominated by the
number of terms in the sum rather than their magnitude: the correlation between
the sum-$\mathrm{CI}_1$ score and the raw path length is Pearson $0.965$ and
Spearman $0.975$ on Erd\H{o}s--R\'enyi graphs ($57{,}402$ back edges pooled over
twenty instances at twelve removal steps each), and Pearson $0.963$ and Spearman
$0.974$ on the configuration model ($56{,}432$ back edges). In practice,
therefore, Cycle-Path CI processes cycles longest-first, and only its final
within-path choice of vertex is influence-based.

The length-normalised variant, Mean-Path CI, was written to test exactly this. If
Cycle-Path CI's advantage came from aggregate cycle influence, dividing by path
length should preserve or improve it. It does not: Mean-Path CI is worse than
Cycle-Path CI on three of the four networks ($0.196$ against $0.192$ on
Erd\H{o}s--R\'enyi, $0.450$ against $0.443$ on Add Health, $0.614$ against
$0.590$ on the CollegeMsg proxy). Only on the configuration model does it come
out ahead, by $0.195$ against $0.196$, and there the two measures disagree:
Mean-Path CI leads marginally on collapse fraction while trailing on area under
the curve ($0.137$ against $0.134$). That one-tenth-of-a-point lead is far
smaller than the standard deviation across the twenty seeds, so it is not a
separation the design can support. Removing the length confound removes whatever
advantage Cycle-Path CI had, which is evidence that the advantage was not coming
from aggregate influence in the first place.

\subsection{Max-Endpoint CI's Gain Is the Ordering, Not the Score}
\label{sec:ordering-not-score}

The same question applies to the rule this paper promotes, and admits the same
kind of control. Max-Endpoint CI differs from Baseline DFS in two ways at once:
it ranks the back edges globally rather than processing the first one the
traversal finds, and it scores endpoints by $\mathrm{CI}_1$ rather than by
degree. A direct comparison of the two cannot say which change earned the
difference. Adopting the global ranking while keeping degree as the score
separates them, and over twenty seeded instances at $n = 500$,
$\langle k \rangle = 3$ the decomposition is one-sided. On Erd\H{o}s--R\'enyi
graphs the ordering is worth $1.07$ percentage points and substituting
$\mathrm{CI}_1$ for degree a further $0.06$, with the influence score ahead on
nine instances of twenty. On the configuration model the ordering is worth $1.65$
points and the influence score is worth $-0.74$: the degree-scored control is the
better rule there, on sixteen instances of twenty.

The supporting correlation agrees. Pooled over every back edge examined, the
maximum endpoint $\mathrm{CI}_1$ tracks the maximum endpoint degree at Pearson
$0.904$ on both families, and the two rules select the same endpoint of the same
back edge on about $90\%$ of back edges. Coreness, the third candidate score
named in Section~\ref{sec:selection}, was measured under the same protocol and is
far worse than either: $0.283$ on both families, eight points behind Baseline
DFS. The reason is structural rather than incidental. At $\langle k \rangle = 3$
these graphs have no $3$-core, so every vertex of the 2-core has coreness exactly
two, on $100\%$ of vertices across all ten instances. A ranking by coreness is
then a complete tie, resolved entirely by traversal order, and the rule
degenerates to processing back edges in discovery order and breaking each at an
arbitrary endpoint.

What Phase~1 contributes is therefore the global cycle ordering. At this scale
the influence measure that motivates it is not separable from degree, on one of
the two families it is worse, and coreness cannot serve as a selection score at
all. This is the sharpest statement of the paper's attribution finding at the
level of a single rule: the named mechanism is not the mechanism that produced
the result.

\subsection{Most of a Synthetic Sequence Is Not the Method}
\label{sec:shared-tail}

Every feedback-vertex-set-based strategy hands off to the identical greedy
$\mathrm{CI}_1$ fill once the graph becomes acyclic, and on the synthetic
families that handoff arrives early. The method-specific portion of each sequence
is between $14.0\%$ and $16.1\%$ of its length; the remaining $83.9\%$ to
$86.0\%$ is produced by a rule shared by all of them. Critically, the collapse
point falls inside that shared tail in all fourteen method-by-graph cells
measured. The synthetic rankings in Tables~\ref{tab:vs-first}
and~\ref{tab:baselines} therefore do not reflect continued method decisions at
the point where they are read off; they reflect the residual graph each method
happened to leave behind at handoff, which the common fill rule then dismantles.
Differences of one or two percentage points between strategies on
Erd\H{o}s--R\'enyi and configuration-model graphs must be interpreted in that
light.

That fraction is not an artifact of the size at which it was measured.
Table~\ref{tab:scaling} repeats the handoff measurement at $n = 500$, $1000$,
$2000$ and $4000$ on both families, over twenty seeded instances at each size.

\begin{table}[htbp]
\centering
\caption{Shared tail and collapse fraction against network size
($\langle k \rangle = 3$, twenty seeded instances per point, seeds 42--61). The
shared-tail column gives the range over all four decycling methods and both
families at that size. ``Best baseline'' is the lowest of CoreHD, Collective
Influence and WEAK-NEIGHBOR.}
\label{tab:scaling}
\setlength{\tabcolsep}{5pt}
\small
\begin{tabular}{ccccccc}
\toprule
 & & \multicolumn{2}{c}{Erd\H{o}s--R\'enyi} & & \multicolumn{2}{c}{Configuration model} \\
$n$ & shared tail & Max-Endpt CI & best baseline & & Max-Endpt CI & best baseline \\
\midrule
500  & $83.9$--$86.1\%$ & \textbf{0.191} & \textbf{0.191} & & 0.196 & \textbf{0.191} \\
1000 & $83.9$--$86.0\%$ & 0.180 & \textbf{0.177} & & 0.180 & \textbf{0.174} \\
2000 & $83.8$--$86.0\%$ & 0.171 & \textbf{0.164} & & 0.172 & \textbf{0.166} \\
4000 & $83.6$--$86.1\%$ & 0.161 & \textbf{0.156} & & 0.164 & \textbf{0.158} \\
\bottomrule
\end{tabular}
\end{table}

The shared tail is remarkably steady: it stays within $83.6\%$ to $86.1\%$ at
every size on both families, moving by less than half a percentage point within
any single method across an eightfold change in $n$. This is the strongest form
in which the finding can be stated, and it is what licenses treating the tail as
a property of the decycling-first family rather than of the operating point at
which it happened to be measured.

What does not survive the same change is the ordering the tail qualifies. Every
collapse fraction falls as $n$ grows, and they do not fall together. At $n = 500$
on Erd\H{o}s--R\'enyi graphs Max-Endpoint CI is level with the best baseline at
$0.191$; by $n = 4000$ it has fallen to $0.161$ while WEAK-NEIGHBOR has fallen to
$0.156$ and CoreHD to $0.160$, leaving it third. On the configuration model it is
behind at every size. The parity reported in Section~\ref{sec:vs-baselines} is
therefore specific to $n$ as well as to $\langle k \rangle$: it holds at five
hundred vertices and dissolves by four thousand. The tail fraction is a stable
property of the family; the ranking it qualifies is not a stable property of the
methods.

The real networks behave differently, and more favourably. There the decycling
phase is long: the method-specific fraction is $42.1\%$ to $45.6\%$ on Add Health
and $56.4\%$ to $61.4\%$ on the CollegeMsg proxy, and the collapse point falls
just past the handoff rather than deep inside the tail. On the real networks,
accordingly, the rankings are genuinely method-driven. The paradox worth noting
is that the regime in which the comparison is most meaningful is also the regime
in which the proposed methods perform least well.

\subsection{Competitive Performance Is Confined to Sparse Graphs}
\label{sec:density}

The narrowing of the field on the real networks is not an artifact of these
implementations but a consequence of the cycle-rank bound of
Lemma~\ref{lem:fvs-density-bound}. Once the average degree reaches four,
$\mu \ge n$, and every feedback vertex set must contain a constant fraction of
the vertices; a decycling-first strategy must then spend that fraction before its
fragmentation phase begins, and the scope for any selection rule to distinguish
itself shrinks in proportion. Table~\ref{tab:density-sweep} measures the
consequence directly, sweeping the average degree from $2.5$ to $6.0$ on both
families at twenty seeded instances per point. The cycle-rank ratio crosses one
at $\langle k \rangle = 4$, exactly where Lemma~\ref{lem:fvs-density-bound}
places the constant-fraction regime, and the fraction of the graph surviving into
the 2-core rises from $0.65$ to $0.98$ across the range.

\begin{table}[htbp]
\centering
\caption{Collapse fraction against average degree ($n = 500$, twenty seeded
instances per point, seeds 42--61). ``Best baseline'' is the lowest of CoreHD,
Collective Influence and WEAK-NEIGHBOR at that point. The cycle-rank column is
the Erd\H{o}s--R\'enyi value; the configuration-model figures agree to within
$0.01$.}
\label{tab:density-sweep}
\setlength{\tabcolsep}{4pt}
\small
\begin{tabular}{ccccccccc}
\toprule
 & & \multicolumn{3}{c}{Erd\H{o}s--R\'enyi} & & \multicolumn{3}{c}{Configuration model} \\
$\langle k \rangle$ & $\mu/n$ & Max-Endpt CI & best baseline & Random & & Max-Endpt CI & best baseline & Random \\
\midrule
2.5 & 0.34 & 0.152 & \textbf{0.145} & 0.617 & & 0.151 & \textbf{0.146} & 0.622 \\
3.0 & 0.55 & \textbf{0.191} & \textbf{0.191} & 0.656 & & 0.196 & \textbf{0.191} & 0.657 \\
3.5 & 0.77 & 0.236 & \textbf{0.229} & 0.696 & & 0.240 & \textbf{0.229} & 0.713 \\
4.0 & 1.00 & 0.281 & \textbf{0.267} & 0.728 & & 0.277 & \textbf{0.264} & 0.738 \\
4.5 & 1.23 & 0.311 & \textbf{0.297} & 0.748 & & 0.309 & \textbf{0.301} & 0.768 \\
5.0 & 1.48 & 0.338 & \textbf{0.329} & 0.775 & & 0.340 & \textbf{0.331} & 0.789 \\
6.0 & 1.97 & 0.397 & \textbf{0.385} & 0.800 & & 0.401 & \textbf{0.386} & 0.798 \\
\bottomrule
\end{tabular}
\end{table}

Two things follow, and the second qualifies the claim this section opened with.
The collapse fraction does rise steadily with density, roughly doubling between
$\langle k \rangle = 2.5$ and $\langle k \rangle = 6$, which is the bound doing
its work. But the structural methods do \emph{not} converge on random removal
over this range: at $\langle k \rangle = 4$, where the bound's constant-fraction
regime begins, they still need $0.267$ against random removal's $0.728$, and at
$\langle k \rangle = 6$ they need $0.385$ against $0.800$. There is no cliff at
$\langle k \rangle = 4$. What the bound predicts is that the budget grows, not
that targeting stops paying, and over this range it continues to pay by a wide
margin.

What does change, and changes early, is the ordering. Max-Endpoint CI's parity
with the established heuristics is confined to $\langle k \rangle = 3$: at $2.5$
it is already behind, and from $3.5$ upward WEAK-NEIGHBOR and CoreHD are ahead of
it at every density on both families. Read with the size dependence of
Section~\ref{sec:shared-tail}, the envelope in which these strategies are
competitive is narrow in two directions at once, an average degree near three and
an $n$ near five hundred, and it is bounded not by the cycle-rank argument but by
the baselines overtaking them.

The two real networks sit far beyond the swept range, at
$\langle k \rangle = 8.24$ with $\mu/n = 3.118$ and $\langle k \rangle = 31.87$
with $\mu/n = 14.936$. The consequence is visible in the spread: on Add Health
the seven structural methods occupy the range $0.427$ to $0.458$, a band of $3.1$
percentage points, while random removal requires $0.829$. This is a structural
constraint on decycling-based dismantling as a family, and it is not addressed by
any refinement of node selection; a method that must remove half the network
before it can begin fragmenting has already spent most of its budget.

\section{Conclusion}
\label{sec:conclusion}

We built a two-phase decycle-then-fragment pipeline, benchmarked it as a method,
and then used it as an instrument to ask what a dismantling comparison in this
family actually reports.

As a method, the result is a parity rather than an advance. Phase~1 returns a
valid feedback vertex set, terminates, and runs in $O(|F|(n+m))$ time, and the
factor of two that appears in the comparison belongs to a local-ratio
subroutine that produces none of the headline numbers. Max-Endpoint CI is level with WEAK-NEIGHBOR on Erd\H{o}s--R\'enyi graphs
at $f^{*} = 0.191$ and ahead of CoreHD's $0.193$, a margin of $0.2$ percentage
points inside instance spread, and it places first on none of the four networks
evaluated. With the fragmentation phase held optimal and identical, so that only
the decycling step varies, it is third of five decyclers on both synthetic
families, within three tenths of a point of the best.

As an instrument, the pipeline produced four measurements that outlast that
parity, and each of them constrains how a margin in this literature should be
read.

\begin{itemize}[itemsep=0.15\baselineskip]
  \item \textbf{Most of a synthetic sequence is shared.} Every
    feedback-vertex-set-based strategy hands off to the same greedy fill, and
    that tail is $83.6\%$ to $86.1\%$ of the sequence at every size from
    $n = 500$ to $n = 4000$, with the collapse point inside it in all fourteen
    measured cells. A synthetic ranking in this family reports the residual each
    method leaves at handoff, not the decisions it keeps making.
  \item \textbf{Fragmentation is not where dismantling is decided.} With
    Braunstein et al.'s exact tree dynamic program substituted for the greedy
    fill, Phase~2 costs $0.019$ to $0.024$ of the network whichever decycler
    precedes it, so $86\%$ to $90\%$ of the removal budget is the feedback vertex
    set itself. Effort spent refining forest fragmentation in this architecture
    is bounded above by about two percentage points.
  \item \textbf{The named mechanism is not always the mechanism.} Max-Endpoint
    CI's gain over the baseline decomposes into a global cycle ordering worth
    $1.07$ and $1.65$ percentage points on the two families, and an influence
    score worth $0.06$ on one and $-0.74$ on the other. Cycle-Path CI's cycle
    score correlates with raw path length at Pearson $0.965$ and $0.963$, so it
    is longest-path-first in practice. Both rules are named for the quantity that
    did not carry the result.
  \item \textbf{The competitive envelope is narrow in two directions.} Parity
    holds at $\langle k \rangle = 3$ and $n = 500$ and is gone by
    $\langle k \rangle = 3.5$ and by $n = 4000$. The cycle-rank bound explains
    why the budget must grow with density, though not a collapse toward random
    removal, which does not occur over the swept range.
\end{itemize}

Two limitations bound what any of this establishes. Real-network results are
single instances and carry no spread, and one of the two networks is a proxy: the
CollegeMsg trim used in place of a proximity network that is not openly
redistributable. And the attribution findings are specific to the decycling-first
template. A shared tail of this size is established for this family, which is
large enough for that to matter; whether it reaches other families of
dismantling method is untested.

The open questions follow from the measurements rather than from the parity.
Since the decycling step carries $86\%$ to $90\%$ of the budget, and since within
that step the cycle ordering rather than the vertex score is what pays, the
promising direction is a decycler that orders cycles by something the dismantling
objective actually rewards, along the lines \citet{li2021longloop} argue for in
distinguishing long loops from short ones. A second direction is to give a
decycler an ordering of its own rather than a set:
Section~\ref{sec:decycling-bound} shows that a set-valued decycler leaves the
area under the curve underdetermined even when its cardinality is bounded. The
third is simply to report the handoff. It costs nothing to measure, it is already
implied by any two-phase method, and without it a reader cannot tell how much of
a published margin belongs to the rule the paper is about.

\subsection*{Reproducibility}

Every number reported here comes from a manifest-carrying run of the accompanying
code at twenty seeds (seeds $42$--$61$); no figure is estimated. The three
selection rules appear in the code under their earlier names,
\texttt{seq\_option1\_max\_ci} (Max-Endpoint CI),
\texttt{seq\_option2\_sum\_ci} (Cycle-Path CI) and
\texttt{seq\_option2\_mean\_ci} (Mean-Path CI). Removal sequences are functions
of the labelled graph and reproduce exactly on a re-run; they are not functions
of the unlabelled structure, since relabelling can change which of several
equally scored candidates is removed.

\bibliography{refs}

\end{document}